\documentclass[preprint,11pt]{elsarticle}
\usepackage[margin=2.5cm]{geometry}
\usepackage{lineno}
\usepackage{subcaption}
\usepackage{adjustbox}
\usepackage{diagbox}
\usepackage{makecell}
\usepackage{booktabs, multirow, array}
\usepackage{cite}

\usepackage{color, soul}

\usepackage[title]{appendix}

\usepackage[justification=centering]{caption}
\usepackage[hidelinks, colorlinks=true,linkcolor=blue]{hyperref}

\begin{document}
	\begin{frontmatter}

		\title{LOD1 3D City Model from LiDAR: The Impact of Segmentation Accuracy on Quality of Urban 3D Modeling and Morphology Extraction}

		\author{Fatemeh Chajaei, Hossein Bagheri}
		\address{Faculty of Civil Engineering and Transportation, University of Isfahan, Isfahan, Iran, h.bagheri@cet.ui.ac.ir}

		\begin{abstract}
		\textcolor{blue}{his is the pre-acceptance version, to read the final version, please go to Remote Sensing Applications: Society and Environment on ScienceDirect, \url{https://www.sciencedirect.com/science/article/abs/pii/S2352938525000874}}. Three-dimensional reconstruction of buildings, particularly at Level of Detail 1 (LOD1), plays a crucial role in various applications such as urban planning, urban environmental studies, and designing optimized transportation networks. This study focuses on assessing the potential of LiDAR data for accurate 3D building reconstruction at LOD1 and extracting morphological features from these models. Four deep semantic segmentation models—U-Net, Attention U-Net, U-Net3+, and DeepLabV3+—were used, applying transfer learning to extract building footprints from LiDAR data. The results showed that U-Net3+ and Attention U-Net outperformed the others, achieving IoU scores of 0.833 and 0.814, respectively. Various statistical measures, including maximum, range, mode, median, and the 90\textsuperscript{th} percentile, were used to estimate building heights, resulting in the generation of 3D models at LOD1. As the main contribution of the research, the impact of segmentation accuracy on the quality of 3D building modeling and the accuracy of morphological features like building area and external wall surface area was investigated. The results showed that the accuracy of building identification (segmentation performance) significantly affects the 3D model quality and the estimation of morphological features, depending on the height calculation method. Overall, the UNet3+ method, utilizing the 90\textsuperscript{th} percentile and median measures, leads to accurate height estimation of buildings and the extraction of morphological features.
		\end{abstract}
		
		\begin{keyword}
			3D building model, LOD1, Semantic segmentation, Deep learning, LiDAR, Urban morphology
		\end{keyword}
	\end{frontmatter}
	
	\section{Introduction}\label{sect_intro}
	The development of Computer-Aided Design (CAD) and Geographic Information Systems (GIS) has significantly contributed to the 3D modeling of urban areas \citep{RN55}. Urban 3D models, which serve as precise digital representations of the terrain and various urban elements such as buildings, trees, vegetation, etc., enable the measurement and extraction of geometric information \citep{RN27}. 
	3D building models have found extensive application across various fields \citep{RN47, RN303}, including urban planning \citep{RN309, RN308, RN302, RN318}, disaster management \citep{RN301, RN300}, navigation systems \citep{RN304, RN305, RN306, RN307}, energy efficiency analysis \citep{RN311, RN313, RN310, RN312, RN314}, environmental impact assessments \citep{RN317, chajaei2024machine}, and visualization and simulations \citep{RN315, RN316, RN319}.

	Various techniques have been employed to create 3D urban models at different levels of detail \citep{RN47}. One of the most commonly used input materials for generating 3D models is Digital Surface Models (DSM), derived from high-resolution remote sensing images \citep{BAGHERI2018389} or LiDAR sensors \citep{RN42}. LiDAR, with its ability to capture high-resolution 3D information over large areas, is particularly effective for creating 3D city models \citep{RN79}. It provides direct georeferenced measurements with suitable planimetric and elevation accuracy and a relatively high point density, making it one of the most significant data acquisition sources for higher-detail 3D building modeling. However, using LiDAR also presents challenges, such as processing large volumes of data, noise appearance, limited ability to represent edges clearly, and the lack of texture mapping for features \citep{RN85}.
	
\begin{figure}[!t]
		\centering
		{\includegraphics[width=\linewidth]{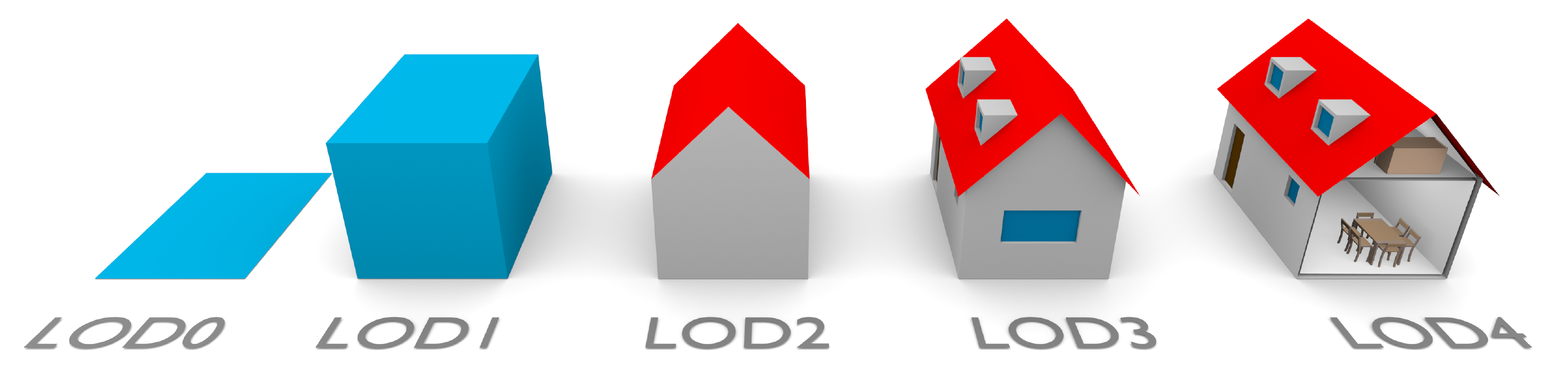}}
		\vspace{0.05cm}
		\caption{CityGML Levels of Detail (LOD0-4) representations \citep{RN130}}
		\label{fig_lod}
	\end{figure}

	3D city models are created with different levels of detail (LOD), which help represent real-world objects in various ways, depending on how complex the model needs to be \citep{RN86}. Representing models at different levels of detail according to users' needs reduces complexity and computational load, allowing for more efficient use of processing power. CityGML, an open standard developed by the Open Geospatial Consortium (OGC) for the representation and storing of 3D city models, categorizes these models into five levels of detail (LOD0 to LOD4), as illustrated in Fig. \ref{fig_lod} \citep{RN87, RN130}. LOD0 represents a 2D model of buildings (footprints) or a polygon of roof surfaces. In LOD1, buildings are depicted as simple blocks with flat roofs, converting the LOD0 model into a 3D model. LOD2 includes simplified roof structures that were added to the model. LOD3 provides a more detailed architectural model featuring precise structures of walls, roofs, balconies, and chimneys. LOD1-3 models enhance the external geometry of the buildings, while LOD4 extends the LOD3 model by adding interior details such as rooms, stairs, and furniture \citep{RN87}.

The accuracy and quality of a 3D building model at the LOD1 depend on the accuracy of building height and footprint parameters. Achieving a more accurate 3D model requires improving the accuracy of these measurements. Various data sources are used to estimate building heights and extract footprints, each with distinct advantages and limitations. Depending on the source of height data, the accuracy of the 3D modeling can vary \citep{RN321, RN320, RN322}. Several studies have focused on improving the height estimation provided by different sensors in the form of DSMs \citep{bagheri2018fusion, bagheri2017uncertainty, bagheri2017fusion, bagheri2018fusionIEEE}. This is also true for building footprints, where higher-resolution data, such as aerial or high-resolution satellite images, lead to more accurate footprint extraction. In addition to the data source, the techniques employed also play a crucial role in the precision of building footprint extraction.

Using satellite or aerial imagery in 3D building modeling requires capturing images with overlapping coverage and generating a DSM through photogrammetry \citep{bagheri2014exploring, bagheri2018exploring}. High-resolution aerial and satellite images are usually more effective on the urban scale due to their wide coverage. However, these images often produce a DSM or point cloud with a resolution of one point per half m\textsuperscript{2}, less dense than LiDAR's. LiDAR point clouds, due to their higher point density, provide a more accurate understanding and better reconstruction of 3D building models. Additionally, the point clouds or DSMs generated through photogrammetric 3D reconstruction techniques tend to have more noise than LiDAR.

Moreover, LiDAR data is inherently georeferenced, providing direct 3D positioning without additional processing, while point clouds from satellite or aerial images require photogrammetric 3D reconstruction techniques that are computationally intensive. Unlike aerial photographs and satellite imagery, LiDAR measurements are unaffected by shadows or relief displacement, ensuring consistent accuracy in varying environmental conditions \citep{RN241}. Despite these, aerial and satellite images typically yield higher accuracy in extracting building footprints. In most previous studies, images have been the primary source of information for extracting building footprints, while LiDAR data has been used as the accurate elevation source for reconstructing 3D building models.

3D building modeling plays a crucial role in extracting morphological parameters of buildings, which are essential for understanding and analyzing the urban structure \citep{RN325, RN326}. Morphological parameters, such as building area, perimeter, height, volume, and facade area, provide quantitative metrics that describe the geometric and physical characteristics of buildings \citep{RN248}. These parameters are widely used in urban planning, resource management, environmental analysis, and disaster risk management. For instance, building height and volume information are critical for shadow analysis, energy efficiency optimization, and earthquake warning systems \citep{RN332, RN327, RN329, RN328, RN330}, while building area and facade area are used to estimate construction costs and evaluate visual impacts \citep{RN331}.

The accuracy of morphological parameters depends heavily on the precision of 3D building models, which relies on the quality of building footprint extraction and height estimation. The existing methods for 3D building modeling face challenges in balancing cost, accuracy, and computational efficiency. Current approaches often rely on a combination of LiDAR data and aerial or satellite imagery. While LiDAR provides high-resolution elevation data, it lacks contextual information such as color and texture, making it challenging to accurately extract building footprints and model complex structures \citep{RN85}. On the other hand, aerial and satellite imagery, though useful for footprint extraction, are computationally intensive, susceptible to environmental factors like shadows, and often require expensive high-resolution data \citep{RN241}. These limitations creates a significant research gap: the need for a cost-effective, scalable, and accurate method for 3D building modeling and morphological parameter extraction that relies solely on LiDAR data. To address this gap, this study proposes developing a LiDAR-exclusive approach that reduces computational overhead by eliminating the need for multi-source data fusion, which typically involves complex data preprocessing, feature alignment, and high storage requirements. By relying solely on LiDAR-derived elevation data, this approach streamlines the modeling workflow, reduces processing time, and enhances scalability while ensuring high accuracy in footprint extraction and height estimation.

Building on this approach, the study develops a method for 3D building modeling at LOD1 using only LiDAR data and extracting morphological parameters with high accuracy. By eliminating the dependency on aerial or satellite imagery, this approach reduces both the financial and computational costs associated with 3D model production, making large-scale urban modeling more feasible. Specifically, the study investigates the feasibility of extracting accurate building footprints directly from LiDAR data and evaluates the impact of footprint accuracy on the reconstruction of 3D buildings. Additionally, it explores the influence of different statistical measures (e.g., maximum, range, mode, median, and 90\textsuperscript{th} percentile) on height estimation and their effect on the accuracy of morphological parameter extraction.

The significance of this research lies in its potential to overcome the limitations of existing methods. By leveraging deep semantic segmentation models to extract building footprints from high-resolution LiDAR-derived DSM (0.23 m) and employing robust statistical measures for height estimation, this study provides a reliable and cost-effective solution for 3D building modeling. Furthermore, the study examines the influence of footprint extraction quality on 3D modeling and the subsequent impact on morphological parameter accuracy, providing valuable insights for improving urban modeling workflows.

This research contributes to the field of urban modeling in several ways. First, it demonstrates the feasibility of using LiDAR data alone for accurate 3D building modeling and morphological parameter extraction, reducing the reliance on costly and computationally intensive imagery. Second, it advances the understanding of how footprint extraction accuracy and height estimation methods impact the quality of 3D models and derived parameters. Finally, the study supports data-driven decision-making in urban planning, sustainability analysis, and smart city development by providing a scalable and cost-effective approach for large-scale 3D modeling. By addressing these challenges, the research not only enhances the accuracy and efficiency of urban modeling but also opens new avenues for automated and large-scale applications in urban analysis and planning.

The findings of this study have wide-ranging practical applications, particularly in urban planning, where accurate 3D models and morphological parameters improve land use, zoning, and infrastructure decisions; energy efficiency analysis, where precise building height and volume data optimize heating, cooling, and lighting systems for sustainability; and disaster management, where reliable 3D models aid in risk assessment, evacuation planning, and recovery efforts. With the rapid growth of smart cities, the integration of artificial intelligence in urban planning, accurate and automated 3D modeling has become an essential tool for data-driven decision-making. As cities continue to expand and face increasing challenges related to climate change, population growth, and resource management, the need for precise, scalable, and cost-effective urban modeling solutions is more urgent than ever. This study addresses these evolving needs by leveraging LiDAR data to provide a cost-effective and scalable approach to 3D building modeling, particularly in areas with limited access to high-resolution imagery. By enhancing the accuracy and efficiency of urban modeling, this research not only supports the development of smarter and more adaptive cities but also contributes to building resilient urban environments capable of addressing future challenges.

This paper is structured into several sections. The introduction and research objectives are provided in the current section. Section \ref{lite} reviews conducted previous studies for 3D building modeling. In the following, Section \ref{sect_data} introduces the study area and describes the data used for extracting building footprints and generating 3D models. Section \ref{sect_methods} describes the techniques for extracting building footprints, reconstructing 3D building models, and extracting morphological parameters. In Section \ref{sect_results}, the results of the segmentation algorithms for extracting building footprints from LiDAR data, as well as the analyses conducted to assess the impact of footprint extraction accuracy on 3D modeling and the extraction of morphological parameters, are presented. Section \ref{sect_discussion} discusses and evaluates the obtained results and various analyses performed.

	\section {Literature Review}\label{lite}
Previous studies have employed various methods and data sources for 3D reconstruction, providing a basis for evaluating the efficiency and accuracy of the techniques used in this research. For example, \citeauthor{RN79} (\citeyear{RN79}) relied solely on LiDAR data for 3D building reconstruction and roof structure modeling. They noted that extracting building information solely from LiDAR is challenging, as most research integrates additional data sources or predefined models. To address this, they identified building blocks directly from the terrain by separating high-elevation building data from low-elevation terrain using Delaunay Triangulation and Voronoi Diagrams \citep{RN79}.

\citeauthor{dlr60168} (\citeyear{dlr60168}) developed a robust approach to generating 3D building models from LiDAR data by integrating processes such as filtering non-ground structures, classifying buildings based on geometric features, and refining building outlines using methods like Minimum Bounding Rectangles (MBR) and Random Sample Consensus (RANSAC). The framework produced models aligned with CityGML standards at various LODs. As noted in the study, since the 1990s, LiDAR data has been preferred over photogrammetry for building extraction due to the high quality of automatically generated DSMs. However, the study highlighted that fully automated reconstruction is challenging for complex structures \citep{dlr60168}.

\citeauthor{RN42} (\citeyear{RN42}) introduced a novel method for automatically generating accurate 3D building models from DSM data and evaluated their proposed method using six different DSM datasets acquired from various sensors in the city of Munich, Germany. Three primary steps comprised the 3D building modeling process: detecting complex building shapes, reconstructing building roofs, and creating 3D visualization. The datasets included three types of DSMs obtained from satellite sensors—Cartosat-1, Ikonos, and WorldView-2—and DSMs derived from airborne sensors, including the 3K camera, High-Resolution Stereo Camera (HRSC), and LiDAR. The performance of the proposed method was assessed based on its ability to identify building footprints, accurately estimate building heights in the model, and the correctness of roof type classification for the different DSM inputs. Additionally, by comparing the outputs, the researchers better understood the capabilities, quality, and applicability of various sensors for 3D modeling. The results indicated that LiDAR DSM visually provided the most accurate 3D model \citep{RN42}.

\citeauthor{RN80} (\citeyear{RN80}) conducted a comparison of various building models created using different input data and methods. The study utilized a range of input data for generating 3D building models at LOD1 and LOD2 levels, including a DSM with 1 m resolution from GeoEye-1 satellite images, a DSM with 50 cm resolution from aerial imagery, a LiDAR-based DSM with 1-m resolution, and topographic maps at 1:1000 and 1:10000 scales. Their findings emphasized the superior accuracy and detail of roof modeling using aerial and LiDAR-derived data compared to satellite imagery-based DSMs. Additionally, the study highlighted that higher accuracy in topographic maps enhances the quality of the resulting building models \citep{RN80}.

\citeauthor{RN36} (\citeyear{RN36}) developed a LOD1 3D model of a part of Ahmedabad, India, using stereo pairs of panchromatic images from Cartosat-1 and high-resolution multispectral imagery from IKONOS. In this study, building boundaries were generated using both automated and semi-automated methods to create the 3D model, and a comparison was made between the models produced by these two approaches \citep{RN36}.

The widespread availability of spatial data, such as vector building boundaries and LiDAR point clouds, has provided opportunities for producing large-scale 3D city models at a lower cost. \citeauthor{RN81} (\citeyear{RN81}) stated that using unclassified point clouds and building boundaries to estimate building heights might give erroneous results due to potential errors in the data. Their study proposed a machine learning-based method for classifying point clouds, which assigned LiDAR points to different classes and finally estimated building heights using only rooftop points. Using a random forest classifier, this method classified LiDAR points into four categories—roof, wall, ground, and outliers. In addition to the point cloud classification, since buildings had different morphologies, they were grouped into four usage types—commercial, residential, skyscrapers, and small buildings—based on a hierarchical clustering algorithm. This grouping was used to assess whether classifying point clouds differently for various building types improved the predictive power of the random forest model. This study aimed to generate LOD1+ models using low-density LiDAR data, which were cheaper and widely available across many world regions, making it computationally efficient for large-scale urban modeling \citep{RN81}.

\citeauthor{RN82} (\citeyear{RN82}) produced LOD3 building models at the city level using a semi-automatically with the highest possible level of automation. For this purpose, they evaluated the usability of various raw data types, including aerial images with a 10 cm Ground Sample Distance (GSD), airborne LiDAR point clouds, terrestrial point clouds from Mobile Mapping systems, and UAV images. The modeling process considered two scenarios: 1) when LOD2 building models were already available and 2) when no prior information about the buildings was provided. The overall process was divided into three main parts, including reconstructing the geometric model of the buildings using images and point clouds, classifying land cover, performing semantic labeling and interactive editing, and modeling the building facades procedurally \citep{RN82}.

\citeauthor{RN96} (\citeyear{RN96}) proposed an accurate and automated method to reconstruct LOD3 building models by integrating multi-source laser point clouds with oblique remote sensing images. The study aimed to extract features from various data sources and use them to create a primary building model, providing a precise reference for automated model editing. To evaluate the proposed algorithm, UAV LiDAR point clouds, terrestrial LiDAR point clouds, and oblique images from an area in Wuhan, China, were utilized. Together, these data sources facilitated a more comprehensive and accurate reconstruction of the LOD3 building model, improving the efficiency and spatial accuracy of the reconstruction process \citep{RN96}.

\citeauthor{RN35} (\citeyear{RN35}) proposed a novel multi-step hybrid (data-driven + model-driven) method for reconstructing LOD2 3D building models using high-resolution WorldView-2 satellite imagery. The roof type is the most important component of the building for reconstruction at LOD2, and to classify it, the study used an image-based method employing a deep learning algorithm. To overcome the low quality of DSM generated from stereo satellite images, auxiliary data such as panchromatic and visually enhanced satellite images were used. In order to evaluate the proposed method, 3D building models were reconstructed in four areas of Munich, and the results were evaluated quantitatively and qualitatively in comparison with LiDAR data as reference data \citep{RN35}.

\citeauthor{baghfusion} (\citeyear{baghfusion}) evaluated the impact of height accuracy on the quality of generating 3D building models at LOD1 using multi-sensor data fusion techniques. They combined building footprints from OSM with height data derived from various sources, including Cartosat-1, TanDEM-X, and SAR-optical stereogrammetry, to create simple 3D models. Their findings showed that data fusion improved the quality of height information, making it feasible to accurately reconstruct 3D buildings over large areas \citep{baghfusion}. 

\citeauthor{RN240} (\citeyear{RN240}) developed a 3D city model for Istanbul, utilizing airborne LiDAR and panoramic imagery to enhance urban planning and analysis. Automatic classification methods were used to process LiDAR data, generate building vectors, and produce digital elevation models (DEM, DSM, nDSM). However, challenges were encountered in distinguishing between buildings, vegetation, and ground surfaces, particularly in densely constructed areas or regions with sloped topography. This led to misclassifications, such as building roofs being categorized as ground or vegetation. High vegetation near buildings further complicated classification. These issues were addressed through manual editing of the LiDAR data, cross-sectional analysis, and corrections informed by aerial photographs, ultimately refining building vectors and improving the accuracy of the classification process \citep{RN240}.

\citeauthor{RN34} (\citeyear{RN34}) introduced a method for creating LOD1 3D city models using very high-resolution stereo satellite images and deep learning algorithms. They employed stereo images from the WorldView-2 satellite. The proposed method first generated a high-resolution multispectral image using pansharpening techniques, and then building boundaries were detected using deep learning algorithms. Ultimately, the building heights were determined using the DSM generated from the stereo images \citep{RN34}.

\citeauthor{rs14092254} (\citeyear{rs14092254}) designed a fully automated method for reconstructing compact 3D building models from large-scale airborne LiDAR point clouds, utilizing available building footprints and an optimized polygonal surface reconstruction framework. They pointed out that a significant challenge in urban reconstruction using airborne LiDAR was building segmentation, as the presence of diverse objects like trees, vehicles, and city furniture in urban scenes makes isolating individual buildings challenging \citep{rs14092254}.

\citeauthor{RN323} (\citeyear{RN323}) proposed a new efficient reconstruction of LOD2 building models from an oblique photogrammetric point cloud guided by façade structures and building footprints. The proposed methodology refined the conventional 3D reconstruction methodologies by abstracted façade structures with regularized monotonic line segments to help build a 2D building topology, which was then extruded to form the 3D model. The findings demonstrated that this method improved the regularity and accuracy of the reconstructed models compared to existing methods like the Polyfit technique, particularly in overcoming noise-related problems and incomplete plane extraction. Although the approach had limitations, such as dependence on the accuracy of building footprints and challenges in representing complex structures, it enabled a feasible solution for large-scale, highly precise, and regular automated production of 3D city models \citep{RN323}.

\citeauthor{RN324} (\citeyear{RN324}) proposed a new framework for effective 3D building model reconstruction from single-view remote sensing images, addressing limitations in traditional multiview image-based methods. Basically, this framework was rooted in a Semantic Flow Field-guided DSM Estimation (SFFDE), fusing global and local features via elevation semantic globalization (ESG) and local-to-global elevation semantic registration (L2G-ESR). This approach significantly enhanced DSM's accuracy, enabling very accurate 3D point cloud and building model reconstructions. Experiments conducted on the ISPRS Vaihingen and DFC2019 datasets demonstrated that 3D building reconstruction provided high accuracy and reduced data acquisition costs, making it an effective method for large-scale urban 3D reconstruction from single-view imagery. \citep{RN324}.

Based on the previous studies, various methods have been developed for reconstructing 3D building models using different data sources and techniques. These methods utilize LiDAR data, satellite images, and aerial imagery and employ image processing algorithms and deep learning techniques to identify buildings and reconstruct their 3D models.

	\section{Study Area and Datasets}\label{sect_data}
	\subsection{Study Area}\label{sub_study}

\begin{figure}[!t]
		\centering
		{\includegraphics[width=\linewidth]{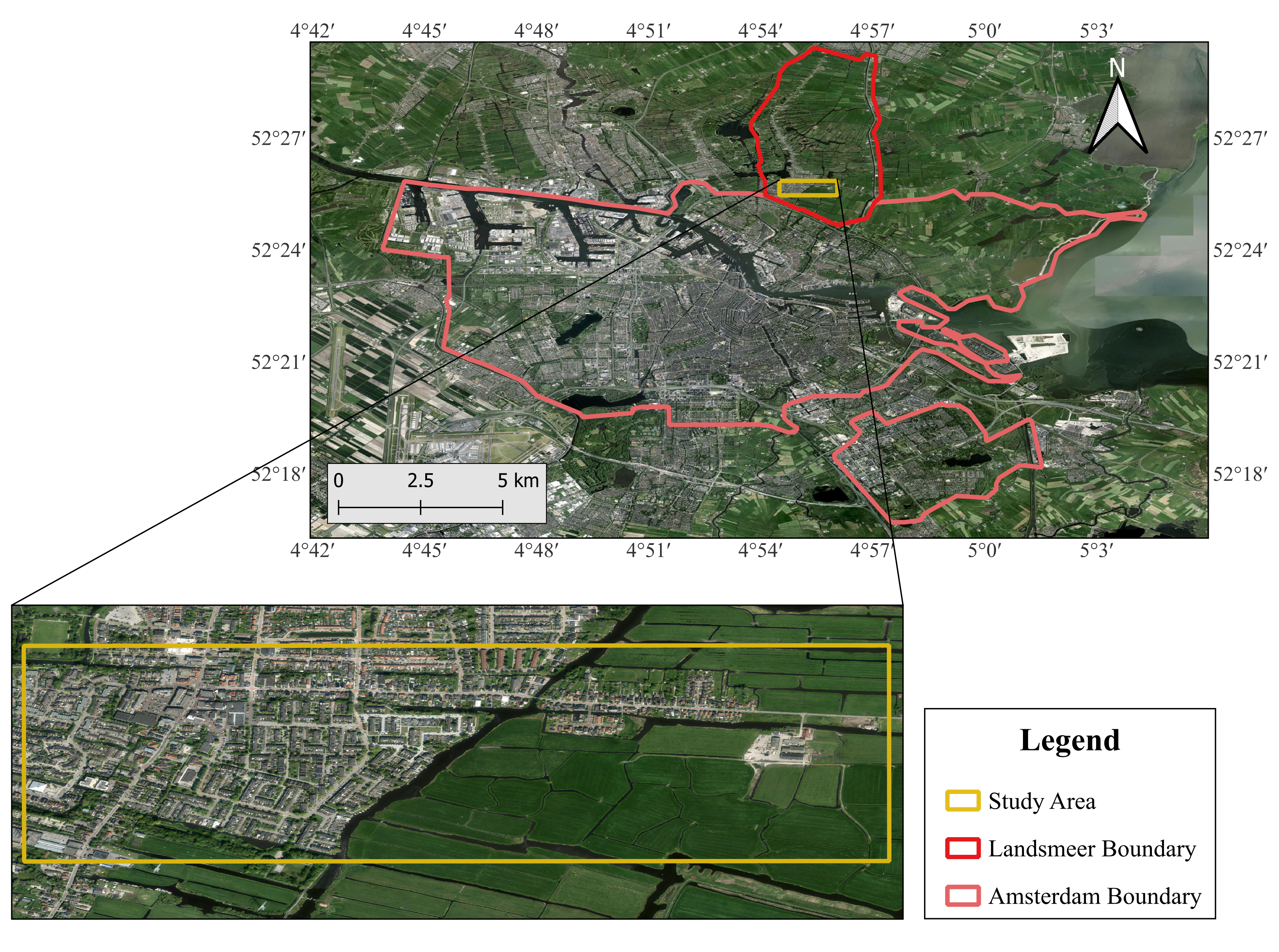}}
		\vspace{0.05cm}
		\caption{A visualization of the study area, part of the Landsmeer, north of Amsterdam.}
		\label{study_area}
	\end{figure}

	This research selected a part of Landsmeer, a city in the province of North Holland in the Netherlands, as the study area. Landsmeer covers an area of approximately 26.5 km\textsuperscript{2} and is located just north of Amsterdam, characterized by its low-lying terrain and distinctive Dutch landscape features. Known for its picturesque canals, green spaces, and suburban environment, Landsmeer provides an interesting case study for exploring the complexities of urban modeling.
Fig. \ref{study_area} visualizes the extent of the study area. 
	
	\subsection{LiDAR Data}\label{sub_lidardata}
	
	Actueel Hoogtebestand Nederland (AHN) provided the data for the study area. The elevation data of the Netherlands (AHN3) was collected between 2014 and 2019. This dataset offers vertical accuracy better than 5 cm, with an average density between 6 and 10 points per m\textsuperscript{2} \citep{RN123}.
	
	Since deep semantic segmentation algorithms require large amounts of training data and the study area lacked accurate building footprint labels,  in addition to including LiDAR data of the study area, data from Miami-Dade County, Florida, USA,  along with the provided labels, was utilized. These data, provided in nDSM format with a spatial resolution of one foot (0.3 m), were employed to train and fine-tune the initial weights of the deep semantic segmentation models for building footprint extraction \citep{RN221}.
	
	\subsection{OpenStreetMap Data}\label{sub_osmdata}

	The building layer from OpenStreetMap (OSM) for Amsterdam and the study area was utilized to create training labels in extracting building footprints. It should be noted that the OSM data used in this study corresponds to the latest update from 2022 \href{https://www.openstreetmap.org/}{(OSM)}. However, the footprints provided by OSM for some buildings are not entirely accurate and require refinement to ensure more precise labeling. These refined labels were then used to fine-tune the deep learning models initially trained on the Miami data.
	
	\section{Methods}\label{sect_methods}
	
\begin{figure}[!t]
		\centering
		{\includegraphics[width=1\linewidth]{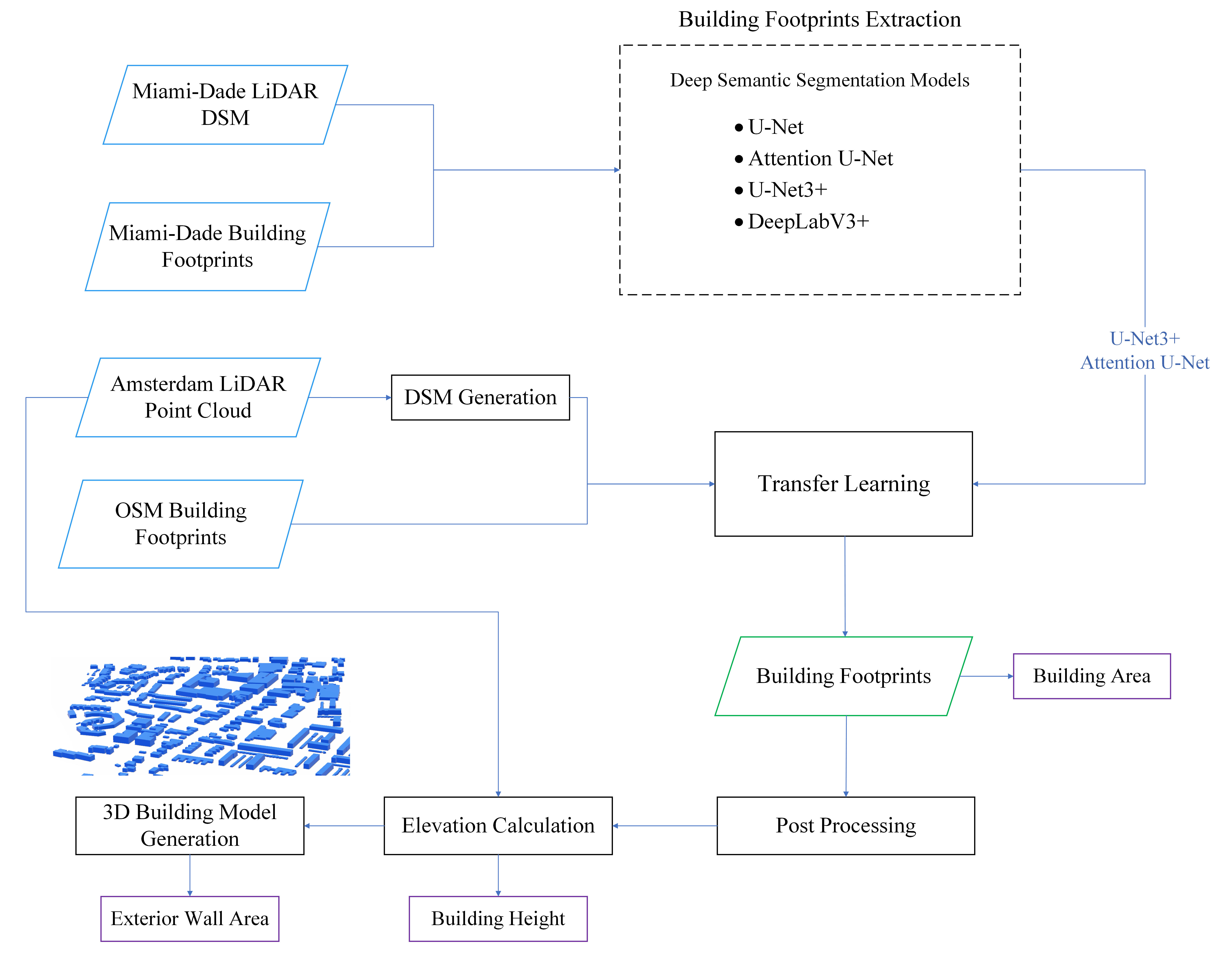}}
		\vspace{0.05cm}
		\caption{The framework used in this study for creating 3D models from LiDAR data and extracting urban morphological parameters.}
		\label{fig_framework}
	\end{figure}

	The first step in this study involves creating accurate 3D models of buildings using LiDAR point clouds. As illustrated in Fig.\ref{fig_framework}, this process includes several vital stages: extracting building footprints using deep semantic segmentation algorithms, using transfer learning techniques to enhance the performance of segmentation models, post-processing of the identified footprints, calculating building heights using different statistical measures, and finally, generating the 3D model. After creating a 3D model, additional morphological parameters are extracted. Each of these stages is explained in more detail in the following sections.
	
	\subsection{LiDAR Data Preparation}\label{sub_methods_lidarprep}
	
	The raw LiDAR data used in this study were initially in the form of a point cloud, containing XYZ coordinates. To make these 3D data usable for analysis, we converted them into DSM format. This conversion was achieved by rasterizing the point cloud using the Adaptive Triangulation method, which preserves elevation details while minimizing interpolation errors. Based on the density and scale of the point cloud data, a grid resolution of 0.23 m was chosen, producing a DSM with a spatial resolution of 0.23 m.

	\subsection{Building Footprint Extraction}\label{sub_methods_bdg}
	To create the 3D model, the building footprints were first extracted from the DSM generated by the LiDAR point cloud using deep learning-based image semantic segmentation algorithms. In this regard, four models, U-Net, Attention U-Net, U-Net3+, and DeepLabV3+, were used, and their performances were compared. The models were trained using the DSM data from the training area and raster label data that distinguished between building and non-building classes, enabling the models to identify building pixels accurately. These models were initially trained on the DSM data of Miami-Dade County, where precise labels were available. The deep semantic segmentation models used in this research are introduced in Appendix \ref{appendix_dl}.

	\subsubsection{Transfer Learning for Improving the Performance of Deep Segmentation Models}\label{subsub_TL}
	
	Transfer learning leverages the knowledge acquired from pre-trained models fully trained on a dataset to solve a similar task \citep{RN129}. Due to the lack of labels (building footprints) in the study area (Amsterdam), the semantic segmentation models were initially trained on the Miami-Dade dataset. However, the structural differences between Miami-Dade and Amsterdam's buildings, along with the limited access to diverse training data, decrease the performance of these models in identifying buildings in Amsterdam. To address this, transfer learning was employed to enhance the model's ability to detect building footprints in the Amsterdam area.
	
	In this approach, deep learning models were first trained on the LiDAR data of Miami-Dade, where accurate training labels were available. Using transfer learning, the model weights were further updated with Amsterdam's data to enhance performance. Models that performed best on the Miami dataset were selected for this weight adjustment process. The LiDAR data and building footprint layers provided by OSM were used as training data for Amsterdam. Although OSM data served as training labels for Amsterdam, these data were not completely accurate or available or correctly aligned spatially for all buildings \citep{RN190, RN147, RN189}. Therefore, only OSM labels that were accurate and matched the LiDAR DSM of the region were selected as training data. Fine-tuning the models with Amsterdam's data improved their generalization ability and led to better performance in the new geographic area.
	
	\subsubsection{Setting up Deep Segmentation Models}\label{subsub_dlsetting}
	
	The model training was configured with 75 epochs to optimize performance over the dataset. The dataset was split into 1,756 samples for training and 311 samples for validation, following an 85\%:15\% split ratio. An additional 422 samples were reserved for the test set to evaluate the model's performance on unseen data. These hyperparameter settings were consistent across all segmentation models.
	
	Table \ref{table_dlsetting} provides an overview of the configurations and architectures of the deep learning models used in this study for identifying building footprints. The models' architectures were designed to reduce validation loss and improve overall performance. To achieve optimal performance, hyperparameter tuning was conducted by experimenting with different learning rates, batch sizes, and variations in pooling and unpooling operations while considering the hardware's computational power. We also tested using backbones; however, increasing the complexity of the model structure led to a decrease in validation accuracy. As a result, no backbone was considered for the U-Net models.
	Notably, as discussed in Appendix \ref{appendix_dl}, the loss function of the U-Net3+ model was different from other models. Data augmentation techniques, including rotation, width and height shift, shear, horizontal flip, and zoom, were employed during training to enhance the generalization capability of the models. The models were trained on an NVIDIA GeForce RTX 3090 GPU, which allowed for efficient processing of the large dataset. The processing time varied between 2 hours 45 minutes and 3 hours 45 minutes, depending on the model architecture, with details specified in Table \ref{table_dlresults}.
	
	A feature aggregation mechanism was applied for the concatenated feature map from five scales in the U-Net3+ model to integrate fine details from shallow layers with high-level semantic information. This mechanism included a Conv2D layer with 160 filters of size 3×3, Batch Normalization, and ReLU activation. In the encoder part of the DeepLabV3+ model, the ASPP module included convolution layers with four dilation rates: 1 (standard convolution), 6, 12, and 18. The extracted features were then passed through a 1×1 convolution layer, linearly upsampled by a factor of 4, and finally concatenated with low-level features of the same spatial resolution from the encoder structure.

	\begin{table}[!t]
		\centering
		\caption{Deep learning model architectures and configurations}
		\label{table_dlsetting}
		\begin{adjustbox}{width=\textwidth}
			\begin{tabular}{c|c|c|c|c}
				\diagbox[width=10em]{Setup}{Models} & U-Net & Attention U-Net & U-Net3+ & DeepLabV3+ \\
				\toprule
				\makecell{Input Dimensions \\ (H x W x D)} & (512, 512, 1) & (512, 512, 1) & (512, 512, 1) & (512, 512, 3) \\
				\midrule
				Number of Filters & [64, 128, 256, 512, 1024] & [64, 128, 256, 512, 1024] & [64, 128, 256, 512, 1024] & Backbone = ResNet50 \\
				\midrule
				\makecell{Activation Functions} & ReLU & ReLU & ReLU & ReLU \\
				\midrule
				\makecell{Output Activation} & Sigmoid & Sigmoid & Sigmoid & Sigmoid \\
				\midrule
				Pool & MaxPooling2D & \makecell{Strided Conv2D \\ + batch norm \\ + activation} & \makecell{Strided Conv2D \\ + batch norm \\ + activation} & ASPP \\
				\midrule
				\makecell{Unpool} & \makecell{Conv2DTranspose \\ + batch norm \\ + activation} & \makecell{Conv2DTranspose \\ + batch norm \\ + activation} & \makecell{Conv2DTranspose \\ + batch norm \\ + activation} & \makecell{2D UpSampling \\ (bilinear)} \\
				\midrule
				Optimizer & Adam (LR: $10^{-4}$) & Adam (LR: $10^{-4}$) & Adam (LR: $10^{-4}$) & Adam (LR: $10^{-4}$) \\
				\midrule
				Loss & Binary Cross Entropy & Binary Cross Entropy & Hybrid Loss & Binary Cross Entropy \\
				\midrule
				Batch Size & 8 & 8 & 4 & 8 \\
			\end{tabular}
		\end{adjustbox}
	\end{table}
	
	\subsubsection{Post-processing}\label{subsub_postprocess}
\begin{figure}[!t]
		\centering
		{\includegraphics[width=0.55\linewidth]{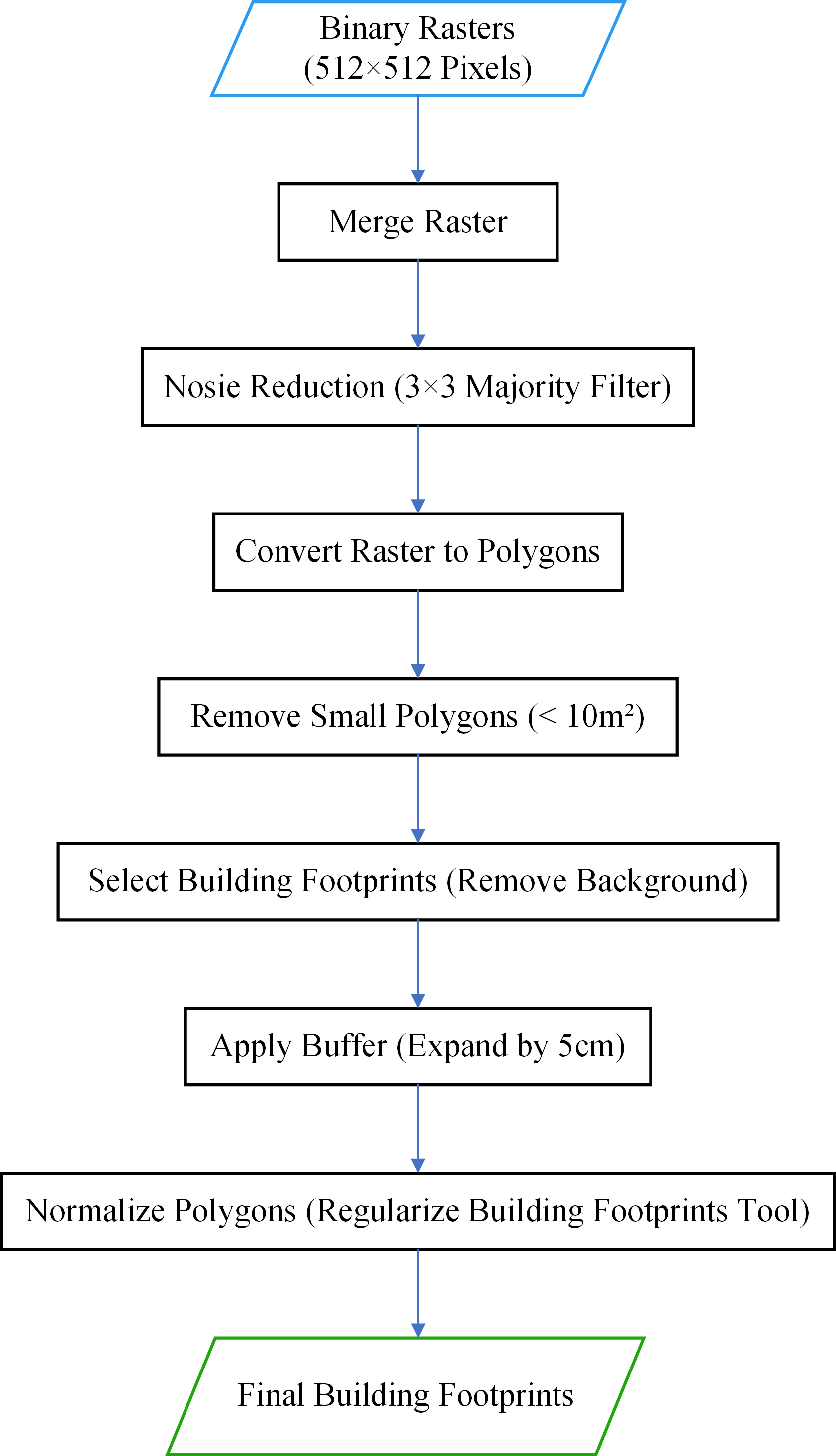}}
		\vspace{0.05cm}
		\caption{Post-processing workflow for refining model outputs and generating building footprints}
		\label{fig_postprocessFlow}
	\end{figure}

	The models' output consists of binary rasters with dimensions of 512×512 pixels, where pixels corresponding to buildings are assigned a value of one, and all other pixels are assigned a zero value. However, these outputs may not be entirely optimal. Therefore, as depicted in Fig.\ref{fig_postprocessFlow}, various post-processing steps are applied to refine the model outputs, correct potential errors, and generate vector files for building footprints. These post-processing steps were carried out using tools available in the ArcGIS Pro environment.
	
	The process begins with merging the 512×512 model outputs to create a single raster. Then, a 3x3 ``Majority Filter'' was applied to reduce noise in the raster. This filter, which replaces each cell based on the majority of its eight neighboring cells, plays a crucial role in refining the model outputs. The building segments in the raster were then converted to polygons, resulting in a vector format. In the next step, tiny areas identified as buildings were removed. For this purpose, polygons with areas below a specified threshold (10 $m^2$) were selected and eliminated.
	
	After selecting and removing background elements, the boundaries of the buildings were expanded by creating a buffer around the polygons by a specific distance (5 cm). Finally, the building polygons were normalized by removing undesirable geometric defects, and the final building footprints were created. This was performed using the ``Regularize Building Footprint'' tool, which employs a polyline compression algorithm to correct undesirable distortions in the building boundaries \citep{RN228}.
	
	\subsection{3D Building Modeling at LOD1}\label{sub_3dmodel}
	
\begin{figure}[!t]
		\centering
		{\includegraphics[width=1\linewidth]{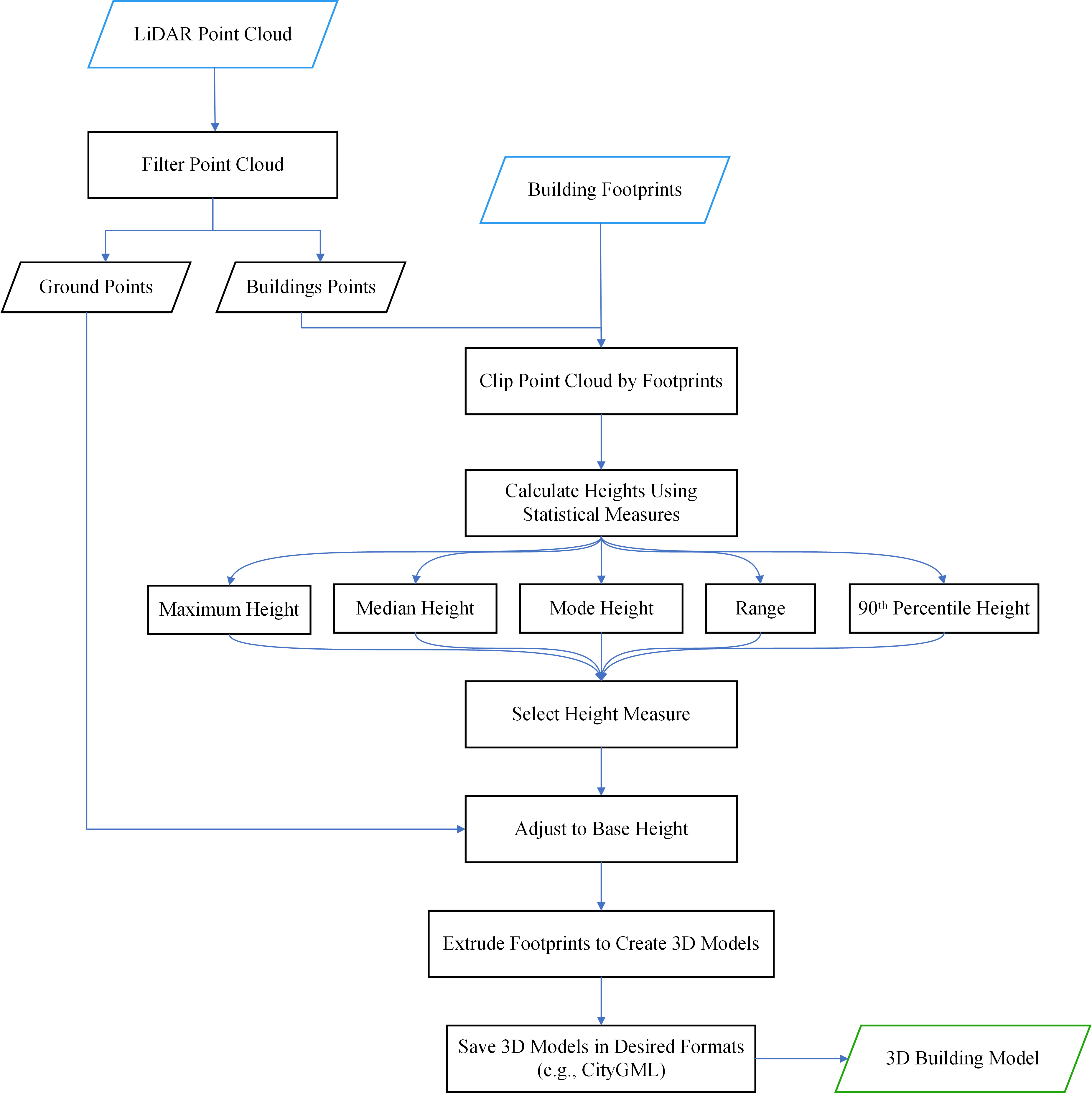}}
		\vspace{0.05cm}
		\caption{Workflow for generating 3D building models at LOD1 using LiDAR data and building footprints}
		\label{fig_3dmodelFlow}
	\end{figure}

	After identifying the building footprints using the aforementioned segmentation models, 3D models of the buildings at LOD1 were generated using height information derived from the LiDAR point cloud. As depicted in Fig.\ref{fig_3dmodelFlow}, the process begins with filtering point cloud data based on classification values to separate building points from other features within the point cloud. The filtered point cloud was then clipped according to the detected footprints.

In the next step, the height of each building was calculated from the LiDAR data. A common approach in height estimation is to calculate the average or median height of all points within a footprint \citep{RN238, RN239}. This study explored various statistical measures, including maximum, median, mode, range, and 90\textsuperscript{th} percentile, each offering distinct perspectives on the height distribution within a building's footprint \citep{RN81, RN237}.

\begin{figure}[!t]
		\centering
		{\includegraphics[width=0.85\linewidth]{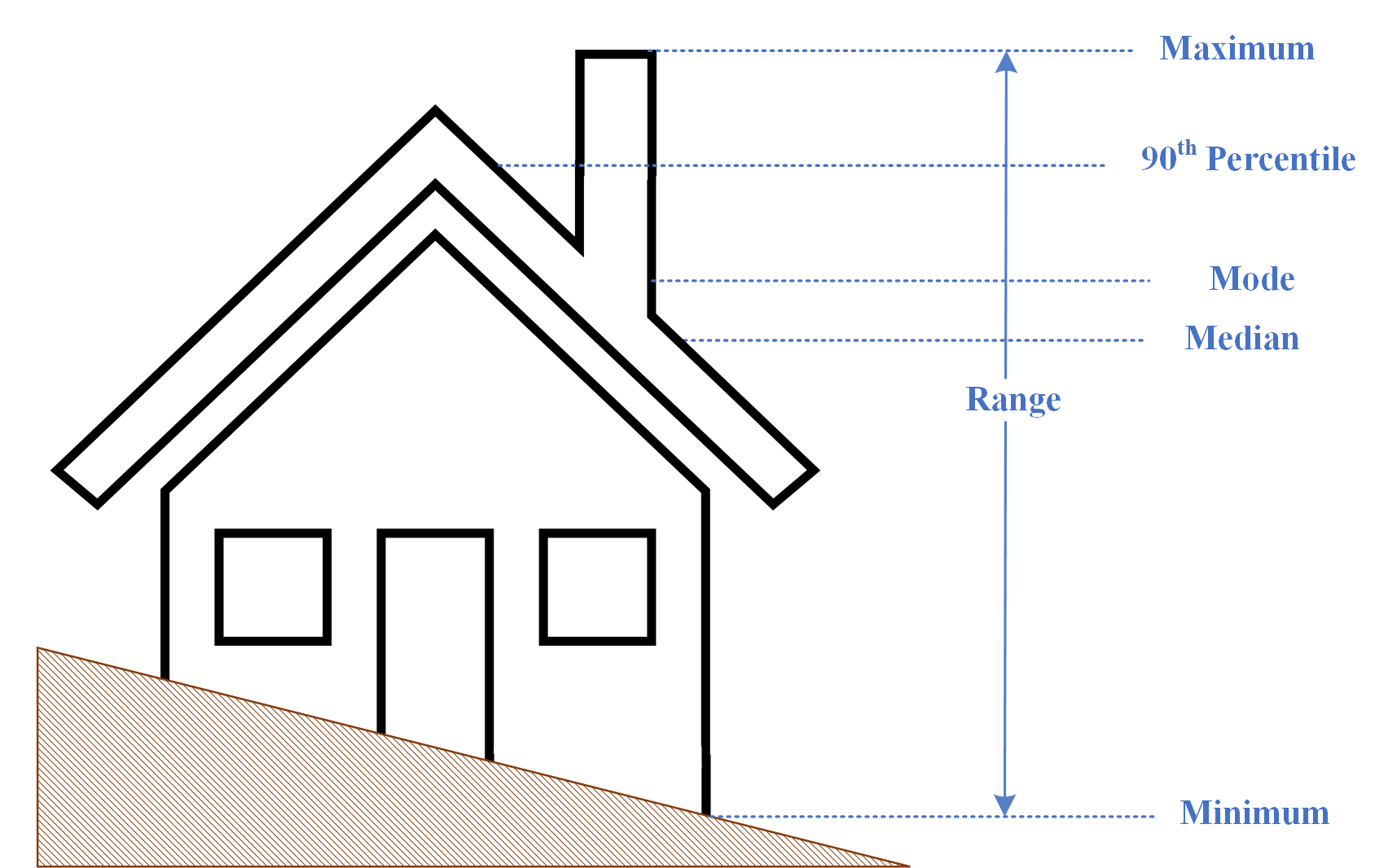}}
		\vspace{0.05cm}
		\caption{Elevation calculation measures for LOD1 building modeling}
		\label{fig_Hmetrics}
	\end{figure}

As illustrated in Fig.\ref{fig_Hmetrics}, the maximum represents the highest elevation point within the building footprint. However, a drawback of using maximum measure is its susceptibility to including elements like antennas, towers, or HVAC equipment that may not be considered part of the building. This metric can introduce outliers into the modeling process, leading to overestimating the building's height. On the other hand, the 90\textsuperscript{th} percentile represents the height value that 90\% of the height data falls under. This measure better reflects the height characteristics of the majority of the building footprint and is less affected by extreme or erroneous values. 

The range is calculated as the difference between the maximum and minimum height points within the building footprint, representing the span of the building's height from its lowest to its highest point. The median is the middle value in the sorted height data (ascending or descending), and the mode indicates the most frequently occurring height value within the building. If there are multiple modes, one of the values is randomly selected.

Using the height values extracted from the LiDAR data, the next step involves transforming the 2D building footprints into 3D space by adjusting their base height. This is achieved by obtaining the base height from the LiDAR ground points within each footprint and applying it to align the building footprints with the ground elevation. Once the base height is set, the final building height is determined based on the aforementioned measures, and the footprints are vertically extruded to generate the 3D building models at LOD1. Finally, the 3D models are saved in the desired output format, such as CityGML.

	\subsection{Building Morphology Extraction}\label{sub_morphology}
	Using the identified building footprints and the generated 3D models, morphological features of the buildings, including the building area and the total area of the exterior walls, were extracted. These features contribute to understanding the structure and spatial characteristics of buildings in the study area.

The building area was calculated using geometric operations on the vector data of the building footprints. The generated 3D models were utilized to calculate the wall area. To achieve this, the building models were first divided into components, including walls, roof, and ground surface. The walls were identified by calculating the slope of each section, with a 90° slope corresponding to walls and a zero slope corresponding to other components. The total wall area was then calculated by summing the surface area of all wall sections for each building.

The results of these calculations were compared across buildings with footprints identified by the U-Net3+ and Attention U-Net models (the best-performing models) and those provided by the OSM data. This comparison was specifically designed to evaluate the compatibility and accuracy of the footprint identification methods and their impact on the extraction of building morphological features.

\subsection{Evaluation Metrics}\label{sub_Hmetrics}
To evaluate the performance of the segmentation models in extracting building footprints from LiDAR data, several metrics, including Intersection over Union (IoU), pixel accuracy, precision, recall, and Dice Coefficient (F1-score), were employed. IoU measures the overlap between the detected footprint and the ground truth \citep{RN131}. Accuracy assesses the correctness of the detected footprints by calculating the percentage of pixels that have been classified correctly \citep{RN113}. At the same time, precision measures the reliability of the correctly classified pixels by calculating the ratio of correctly detected footprints to the total number of pixels recognized as buildings \citep{RN132}. Recall evaluates the completeness of the positive predictions by determining the ratio of correctly predicted footprints to the total ground truth footprints. The Dice Coefficient combines precision and recall and provides a balanced evaluation metric for assessing segmentation performance.

The main goal of this study is to assess how the accuracy of building footprint extraction impacts the precision of 3D building models and the quality of their morphological features. To achieve this, the accuracy of 3D reconstructions created using different segmentation methods was evaluated by comparing building height estimates with reference data (such as building footprints from OpenStreetMap) for each building. For this purpose, accuracy analysis was conducted using three widely adopted statistical measures: RMSE, MAE, and R². RMSE was selected for its ability to quantify overall error magnitude while giving greater weight to larger deviations, making it particularly useful for identifying outliers in height estimation. MAE provides a direct interpretation of average absolute error, making it useful for understanding general deviations from reference heights. R² evaluates how well the estimated heights align with reference data. Together, they enable a robust assessment of both absolute and relative accuracy, ensuring a comprehensive validation of height estimation methods. Similarly, the impact of the quality of footprints extracted by different methods on the accuracy of derived morphological features was examined. In this case, OSM data were also utilized as the reference.

These metrics provide a comprehensive assessment of segmentation results' accuracy and performance, as well as their impact on building height estimation using different statistical measures. They also offer insights into the models' strengths and limitations, allowing for comparisons between different approaches or techniques employed in the study.
	
	\section{Results}\label{sect_results}
	\subsection{Performance of Deep Segmentation Models for Building Footprint Extraction}\label{sub_dlResults}

	\begin{table}[!t]
	\centering
	\caption{Performance Evaluation of Deep Segmentation Models on the Miami-Dade Dataset with 95\% Confidence Intervals (t-Distribution, SEM)}
	\label{table_dlresults}
	\begin{adjustbox}{width=\textwidth}
		\begin{tabular}{c|c|c|c|c|c|c}
			Model & IoU & Dice Coefficient & Pixel Accuracy & {Precision} & {Recall} & Training Time \\ 
			\toprule
			{U-Net} & 0.841 ± 0.012 & 0.913 ± 0.010 & 0.967 ± 0.002 & 0.899 ± 0.012 & 0.928 ± 0.008 & 3\textsuperscript{h}21\textsuperscript{m} \\ 
			\midrule
			{Attention U-Net} & 0.852 ± 0.012 & 0.920 ± 0.010 & \textbf{0.970 ± 0.002} & \textbf{0.908 ± 0.012} & 0.933 ± 0.008 & \textbf{2\textsuperscript{h}44\textsuperscript{m}} \\ 
			\midrule
			{U-Net3+} & \textbf{0.853 ± 0.012} & \textbf{0.921 ± 0.010} & \textbf{0.970 ± 0.002} & 0.904 ± 0.011 & \textbf{0.938 ± 0.008} & 3\textsuperscript{h}09\textsuperscript{m} \\ 
			\midrule
			{DeepLabV3+} & 0.842 ± 0.012 & 0.914 ± 0.010 & 0.968 ± 0.002 & 0.907 ± 0.010 & 0.922 ± 0.010 & 3\textsuperscript{h}42\textsuperscript{m} \\ 
		\end{tabular}
	\end{adjustbox}
\end{table}
	To create 3D models of buildings, deep semantic segmentation models were initially used to extract building footprints. The performance of deep learning models, including U-Net, Attention U-Net, U-Net3+, and DeepLabV3+, was evaluated for identifying building footprints from the Miami-Dade test dataset. The results, summarized in Table \ref{table_dlresults}, including 95\% confidence intervals for each evaluation metric, were computed using the t-distribution and standard error of the mean (SEM) to measure statistical reliability.
	
	The findings indicate indicate that all models achieved acceptable and nearly similar performance across various evaluation metrics. However, the U-Net3+ model performed the best, closely followed by the Attention U-Net, which produced results similar to U-Net3+ regarding building footprint extraction accuracy. Therefore, these two models were selected as the top-performing models for applying segmentation and extracting building footprints in the study area.

It is worth noting that the Attention U-Net model demonstrated the shortest training time among all models, making it an efficient choice for scenarios with limited computational resources. The U-Net3+ model ranked next with a slight time difference. Due to computational constraints, the U-Net3+ model was trained with a smaller batch size (4) than the other models (8), which may explain the longer training time relative to the Attention U-Net model.
	
	\begin{table}[!t]
		\centering
		\caption{Performance evaluation of deep segmentation models on study area dataset before applying transfer learning}
		\label{table_befTL}
		\begin{adjustbox}{width=0.8\textwidth}
			\begin{tabular}{c|c|c|c|c|c}
				Model & IoU & Dice Coefficient & Pixel Accuracy & Precision & Recall \\ 
				\toprule
				U-Net3+ & \textbf{0.597} & \textbf{0.747} & \textbf{0.906} & 0.937 & \textbf{0.622} \\  
				\midrule
				Attention U-Net & 0.557 & 0.716 & 0.897 & \textbf{0.943} & 0.577 \\ 
			\end{tabular}
		\end{adjustbox}
	\end{table}
	
	\begin{table}[!t]
		\centering
		\caption{Performance evaluation of deep segmentation models on study area dataset after applying transfer learning}
		\label{table_aftTL}
		\begin{adjustbox}{width=0.8\textwidth}
			\begin{tabular}{c|c|c|c|c|c}
				Model & IoU & Dice Coefficient & Pixel Accuracy & Precision & Recall \\ 
				\toprule
				U-Net3+ & \textbf{0.833} & \textbf{0.909} & \textbf{0.961} & \textbf{0.945} & \textbf{0.875} \\ 
				\midrule
				Attention U-Net & 0.814 & 0.898 & 0.956 & 0.943 & 0.857 \\  
			\end{tabular}
		\end{adjustbox}
	\end{table}
	
	Although these models demonstrated promising results in the Miami-Dade test areas, as previously mentioned, the capability of these models to identify footprints in the study area decreased due to limited access to training data and structural differences between buildings in the two cities. The top two models, U-Net3+ and Attention U-Net, were refined through transfer learning by updating their weights to overcome this limitation. 
	
	Table \ref{table_befTL} shows the evaluation results of these two models for the study area's test data without applying transfer learning. In contrast, Table \ref{table_aftTL} shows the evaluation results after applying transfer learning. For example, the U-Net3+ model achieved an IoU of 0.853 on the Miami-Dade test data, but the IoU for the study area dropped to 0.597. However, after using the transfer learning technique, the IoU for the study area improved to 0.833. 
	
	The results indicate a significant improvement in the performance of both U-Net3+ and Attention U-Net after applying transfer learning. This improvement, highlighted by the U-Net3+ model achieving an IoU of 0.833 and a Dice Coefficient of 0.909 and the Attention U-Net model reaching an IoU of 0.814 and a Dice Coefficient of 0.898, is indeed impressive. The precision of both models was very similar (U-Net3+ with 0.9453 and Attention U-Net with 0.9431), indicating both models' high capability to identify building footprints accurately. However, based on the metrics, the U-Net3+ model showed slightly better overall performance, further impressing its results.
	
	Overall, these results demonstrate the effectiveness of transfer learning in adapting deep learning models trained in one geographical area to another with different characteristics. Consequently, using transfer learning improved the performance of building footprint detection models on study area data. These findings underscore the potential of transfer learning as a valuable technique to enhance the performance of deep learning models, especially in scenarios where access to diverse and labeled training data is limited.

	\begin{table}[!t]
		\centering
		\caption{Comparison of quality of 3D building models generated using various measures and footprints identified by the Attention U-Net and U-Net3+ models.}
		\label{table_HeightError}
		\begin{adjustbox}{width=\textwidth}
			\begin{tabular}{c|c|c|c|c|c|c}
				\multirow{2}{*}{Z} & \multicolumn{2}{c|}{RMSE (m)} & \multicolumn{2}{c|}{MAE (m)} & \multicolumn{2}{c}{R\textsuperscript{2}} \\
				\cline{2-7}
				& \rule{0pt}{3ex} U-Net3+ & Attention U-Net & U-Net3+ & Attention U-Net & U-Net3+ & Attention U-Net \\ 
				\toprule
				Median & 0.564 & 0.544 & 0.208 & 0.219 & 0.919 & 0.924 \\
				\midrule
				Max & 0.736 & 0.765 & 0.121 & 0.161 & 0.936 & 0.931 \\
				\midrule
				Mode & 1.168 & 1.197 & 0.397 & 0.435 & 0.704 & 0.689 \\
				\midrule
				Range & 1.151 & 1.146 & 0.406 & 0.447 & 0.854 & 0.855 \\
				\midrule
				90\textsuperscript{th} Percentile & 0.622 & 0.555 & 0.116 & 0.127 & 0.936 & 0.949 \\
			\end{tabular}
		\end{adjustbox}
	\end{table}
	
	Table \ref{table_HeightError} shows the impact of the accuracy of building footprint extraction (the performance of different segmentation models) on the 3D modeling of buildings, considering various height reconstruction measures. For this evaluation, the building reconstruction process was also carried out using OSM footprint data as a reference. The 3D models generated from different footprints were then compared with those reconstructed from OSM footprints based on various metrics, including RMSE, MAE, and R², to provide a comprehensive insight into the quality of the 3D models generated based on different height calculation measures. The results of this comparison are presented in Table \ref{table_HeightError}. As mentioned in Section \ref{sub_3dmodel}, the height of each building was estimated considering metrics such as median, maximum, mode, range, and the 90\textsuperscript{th} percentile derived from LiDAR point cloud data. Overall, the evaluation results show that both the Attention U-Net and U-Net3+ models perform well in estimating building heights. However, the U-Net3+ model performs relatively better in reconstructing the LOD1 model based on various height measures.
	
	Additionally, the height estimation error is minimized when using the median and 90\textsuperscript{th} percentile metrics, whereas height estimation using mode and range metrics shows the highest errors. These results suggest that the uncertainty in building footprint extraction has less impact on 3D reconstruction based on the median and 90\textsuperscript{th} percentile measures. In contrast, it has the most significant impact on 3D model reconstruction using other measures.
	
	Statistical measures such as mode and maximum can be influenced by skewed distributions or heterogeneous data distribution. As a result, variations in the distribution of building points due to uncertainty in footprint extraction can lead to differing height estimates. In contrast, the median and the 90\textsuperscript{th} percentile, less susceptible to such biases, provide more reliable estimates, even in non-normal distributions. This robustness of the median and 90\textsuperscript{th} percentile measures, even in the face of data distribution variations, should provide a sense of reassurance about their reliability compared to measures like mode and range.
	
	Height calculation using the mode measure represents the most common LiDAR height within the corresponding building footprint. However, in cases where data has an uneven distribution, the mode metric may produce inaccurate results, leading to increased errors and reduced accuracy in height estimation. For instance, in buildings with sloped roofs, even slight changes in building footprints can result in heterogeneous height distributions, altering the frequency of height values and leading to more significant errors in final height estimation. Therefore, in scenarios with significant variations in heights and roofs with varying slopes, using the mode measure for height calculation may cause more errors due to the accuracy of footprint detection. As a result, the mode is not a stable measure for final height estimation and is highly dependent on the building footprints.

	\begin{figure}[!t]
		\centering
		\includegraphics[width=\linewidth]{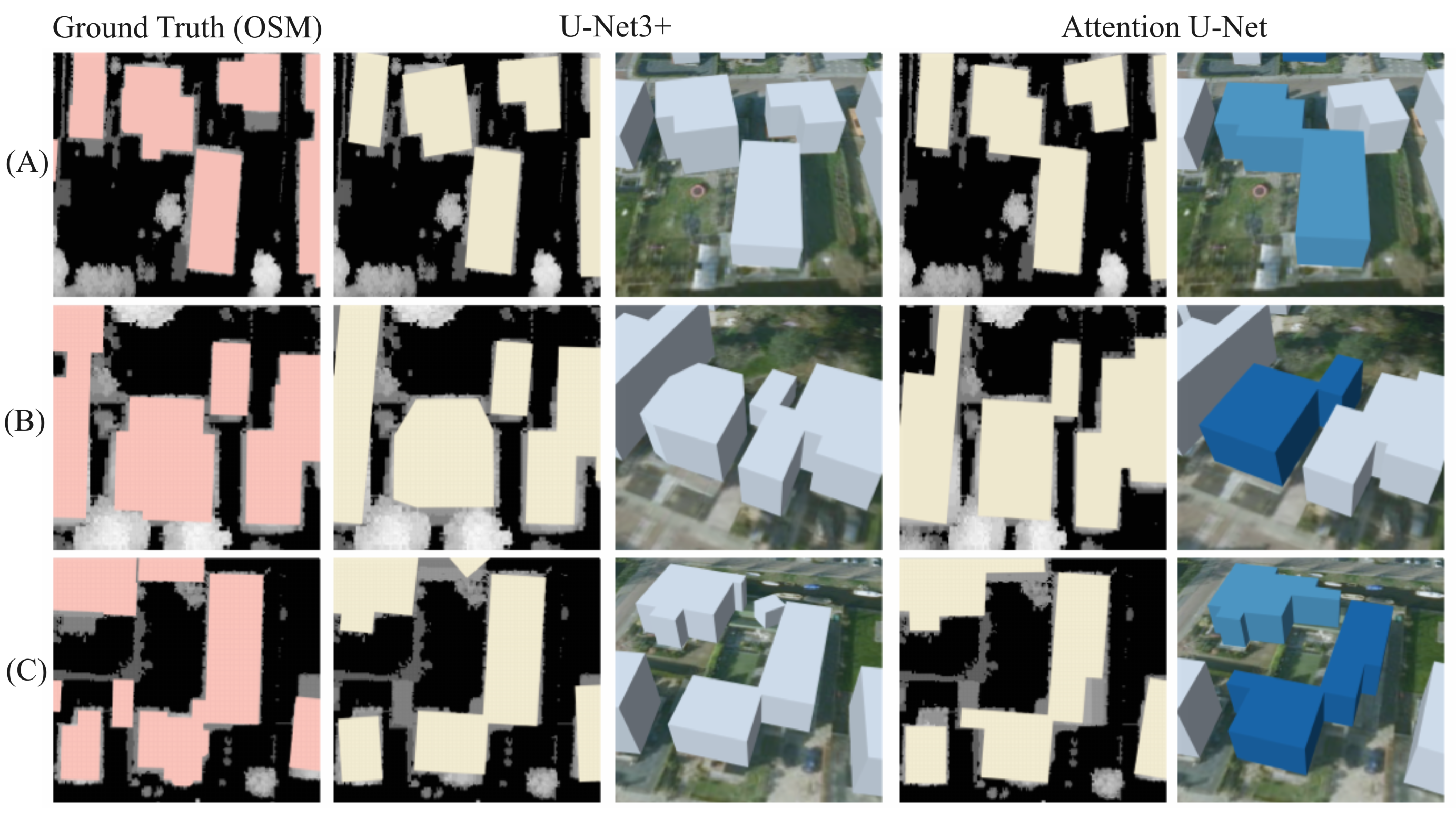}
		\vspace{0.2cm}
		
		\includegraphics[width=0.9\linewidth]{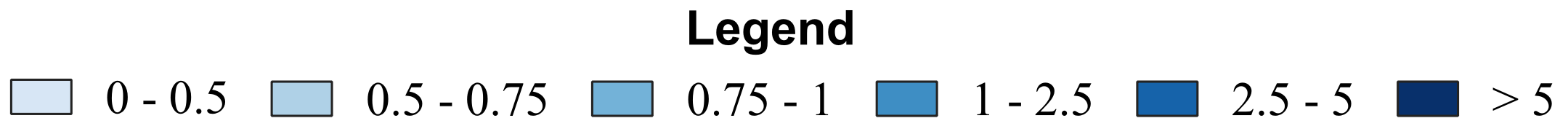}
		\vspace{0.05cm}
		
		\caption{Impact of uncertainties in building footprint detection by different segmentation models on estimating elevations of buildings}
		\label{fig_HbdgErr}
	\end{figure}
	
	Fig.\ref{fig_HbdgErr} displays the ground truth footprints (OSM) alongside the footprints identified using the U-Net3+ and Attention U-Net models for several sample buildings. Also, the corresponding LOD1 model is shown for each footprint based on the 90\textsuperscript{th} percentile height measure. Additionally, the impact of 3D model reconstruction error due to uncertainty in footprint detection by the segmentation models is visually represented using color coding. This visual analysis provides valuable insights into the models' performance in identifying building footprints and estimating their heights. As shown in the figure, the U-Net3+ model exhibits superior performance compared to the Attention U-Net model in accurately detecting separate footprints, resulting in less error in the height estimation of the 3D model. Specifically, in cases A and B in Fig.\ref{fig_HbdgErr}, the Attention U-Net model struggles to distinguish between closely situated buildings, sometimes merging adjacent footprints and recognizing them as a single building. This misinterpretation leads to higher height estimation errors.
	
	\subsection{Result of Building Morphology Extraction}\label{sub_MorphologyResults}

	This section presents the impact of uncertainty in identifying building footprints on estimating morphological features, such as footprint area and the total exterior wall area. Initially, the difference in calculated area between the footprints identified using the two deep learning models (U-Net3+ and Attention U-Net) and the reference OSM data was computed. Similarly, the discrepancy in the exterior wall area resulting from the 3D models reconstructed using the footprints identified by the U-Net3+ and Attention U-Net models, based on the median and 90\textsuperscript{th} percentile height measures, compared to the models derived from OSM data, was calculated.

	Table \ref{table_areaEval} presents the area estimation error for the Attention U-Net and U-Net3+ models based on different evaluation metrics, including RMSE, MAE, and  R². The correlation (R²) between the building areas extracted using the two deep learning models and the reference areas provided by OSM is high (R² = 0.96), indicating that both segmentation models perform well in detecting building footprints. Despite the similar R² values for both models, the higher RMSE value for the Attention U-Net model (79.37) compared to U-Net3+ (74.59), and correspondingly, the higher MAE value for Attention U-Net (35.66) compared to U-Net3+ (32.81), suggest that the U-Net3+ model performs better in extracting area morphology. These results suggest that while both models are effective, U-Net3+ is more accurate in extracting building areas, with the Attention U-Net model producing relatively higher area estimation errors.
	
	Similarly, Table \ref{table_wallEval} compares the impact of uncertainty in building footprints derived from the Attention U-Net and U-Net3+ models on estimating the total surface area of building exterior walls. The table includes RMSE, MAE, and R² values for wall area estimation using two different height measures: median and 90\textsuperscript{th} percentile. The results indicate that, for both height measures, the U-Net3+ model yields lower errors in terms of RMSE and MAE, suggesting better performance in estimating wall areas compared to the Attention U-Net model. However, both segmentation models show higher errors when using the 90\textsuperscript{th} percentile height metric. This is likely due to the 90\textsuperscript{th} percentile focusing on the upper range of the height distribution, causing overestimation of building heights for structures with varying and non-uniform elevations. These overestimations lead to larger wall surface areas and, subsequently, higher RMSE and MAE values.
	
\begin{figure}[!t]
		\centering
		{\includegraphics[width=\linewidth]{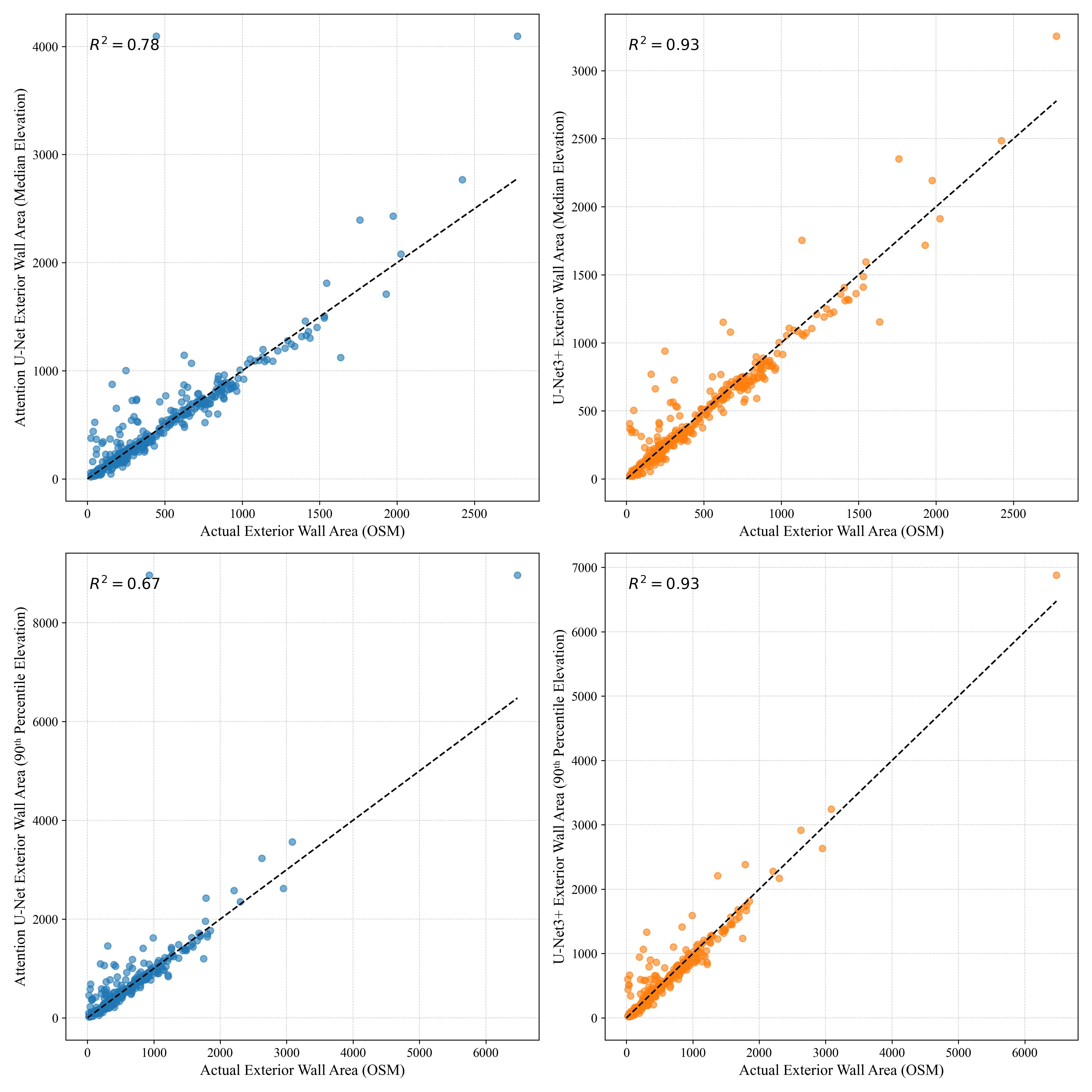}}
		\vspace{0.05cm}
		\caption{Correlation plots comparing estimated and actual wall surface areas using footprints obtained from segmentation models against OSM data, considering different height measures.}
		\label{fig_wallEval_R2}
	\end{figure}

	Additionally, the R² values shown in Fig.\ref{fig_wallEval_R2} demonstrate the correlation between the actual and estimated wall areas derived from 3D models. For the Attention U-Net model, R² values are 0.78 and 0.67 for the median and 90\textsuperscript{th} percentile height measures, respectively. In contrast, the U-Net3+ model shows higher R² values of 0.93 for both height measures, indicating its superior performance in estimating wall areas across different height estimation measures.
	
	Overall, the findings reveal that the quality of building footprint extraction and the choice of height metric significantly impact the accuracy of morphological parameter calculations, such as area and exterior wall surface. Therefore, the need for more precise extraction of building footprints is urgent, as it is essential for achieving high accuracy in morphology calculations.

	\begin{table}[!t]
		\centering
		\caption{Accuracy of estimated building areas extracted by the U-Net3+ and Attention U-Net models}
		\label{table_areaEval}
			\begin{tabular}{c|c|c|c}
				Model & RMSE (m²) & MAE (m²) & R² \\ 
				\toprule
				U-Net3+ & 74.59  & 32.81  & 0.96  \\ 
				\midrule
				Attention U-Net & 79.37 & 35.66 & 0.96 \\  
			\end{tabular}
	\end{table}

	\begin{table}[!t]
		\centering
		\caption{Comparison of various metrics for wall surface area estimation using footprints obtained from segmentation models against OSM data, considering different height measures.}
		\label{table_wallEval}
		\begin{adjustbox}{width=\textwidth}
			\begin{tabular}{c|c|c|c|c|c|c}
				\multirow{2}{*}{Model} & \multicolumn{2}{c|}{RMSE (m)} & \multicolumn{2}{c|}{MAE (m)} & \multicolumn{2}{c}{R\textsuperscript{2}} \\
				\cline{2-7}
				& \rule{0pt}{3ex} Median & 90\textsuperscript{th} Percentile & Median & 90\textsuperscript{th} Percentile & Median & 90\textsuperscript{th} Percentile \\ 
				\toprule
				U-Net3+ & 103.84 & 138.23 & 49.34 & 65.92 & 0.93 & 0.93 \\
				\midrule
				Attention U-Net & 206.95 & 409.09 & 62.4 & 92.02 & 0.78 & 0.67 \\
			\end{tabular}
		\end{adjustbox}
	\end{table}

	\section{Discussion}\label{sect_discussion}
	\subsection{Finding Overview}\label{sub_discuss_finding}

	This study investigated the feasibility of generating LOD1 3D building models using only LiDAR data, addressing three core research questions: (1) How does the accuracy of building footprint extraction impact the quality of 3D building reconstruction? (2) Which statistical measures for height estimation (e.g., maximum, range, mode, median, and 90\textsuperscript{th} percentile) yield the most accurate results for LOD1 modeling? and (3) How does the quality of footprint extraction and height estimation influence the accuracy of derived morphological parameters, such as building area and the total area of exterior walls?
	
	Our findings highlight the critical role of segmentation performance in determining footprint accuracy, which directly affects height estimation and the overall quality of 3D models. Fig. 9 summarizes the key findings, demonstrating how segmentation accuracy, height estimation methods, and footprint quality collectively influence the reliability of 3D building modeling and urban morphology analysis. The following sections provide a detailed discussion of these findings, examining the interplay between segmentation accuracy, height estimation techniques, and footprint quality and their implications for improving urban modeling workflows.
	
	\begin{figure}[!t]
		\centering
		{\includegraphics[width=\linewidth]{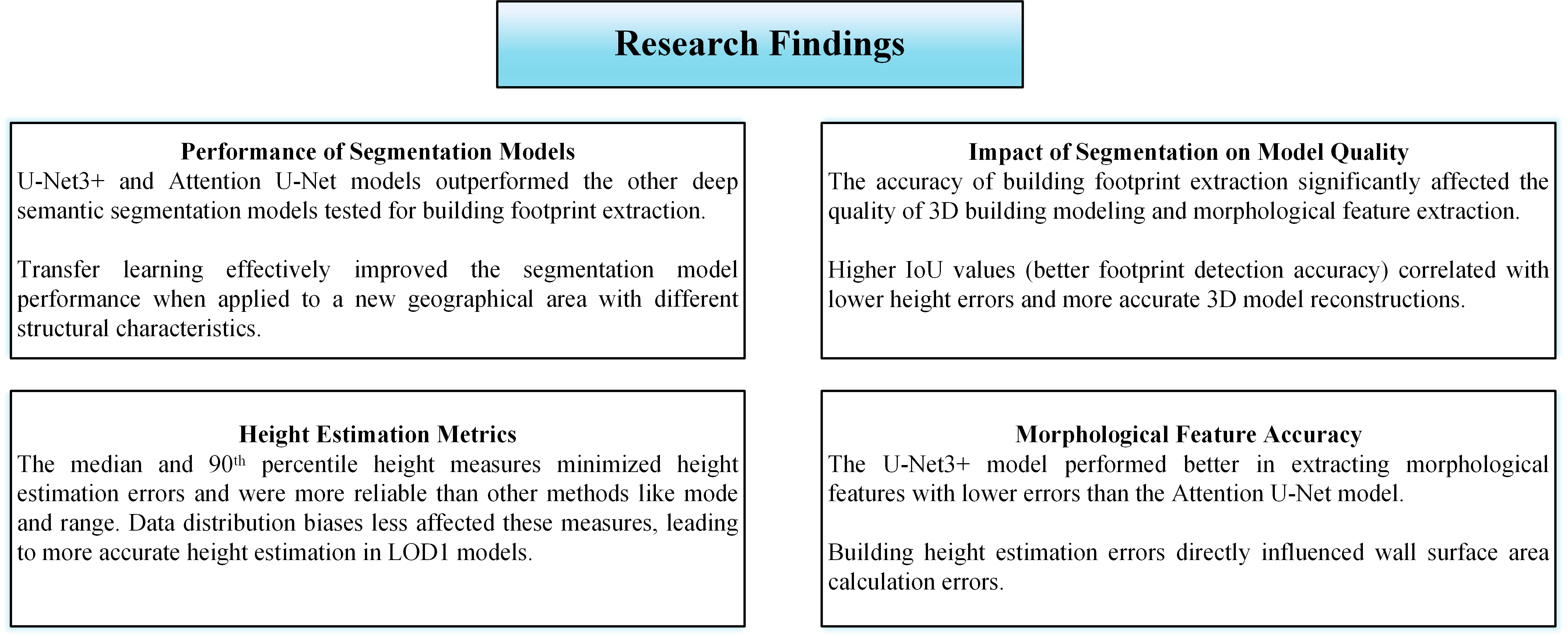}}
		\vspace{0.05cm}
		\caption{Summary of key research findings}
		\label{fig_finding}
	\end{figure}
	
	\subsubsection{Impact of Segmentation Performance on the Accuracy of 3D Modeling}\label{sub_3dAccDiscuss}
	
	\begin{figure}[!t]
		\centering
		{\includegraphics[width=\linewidth]{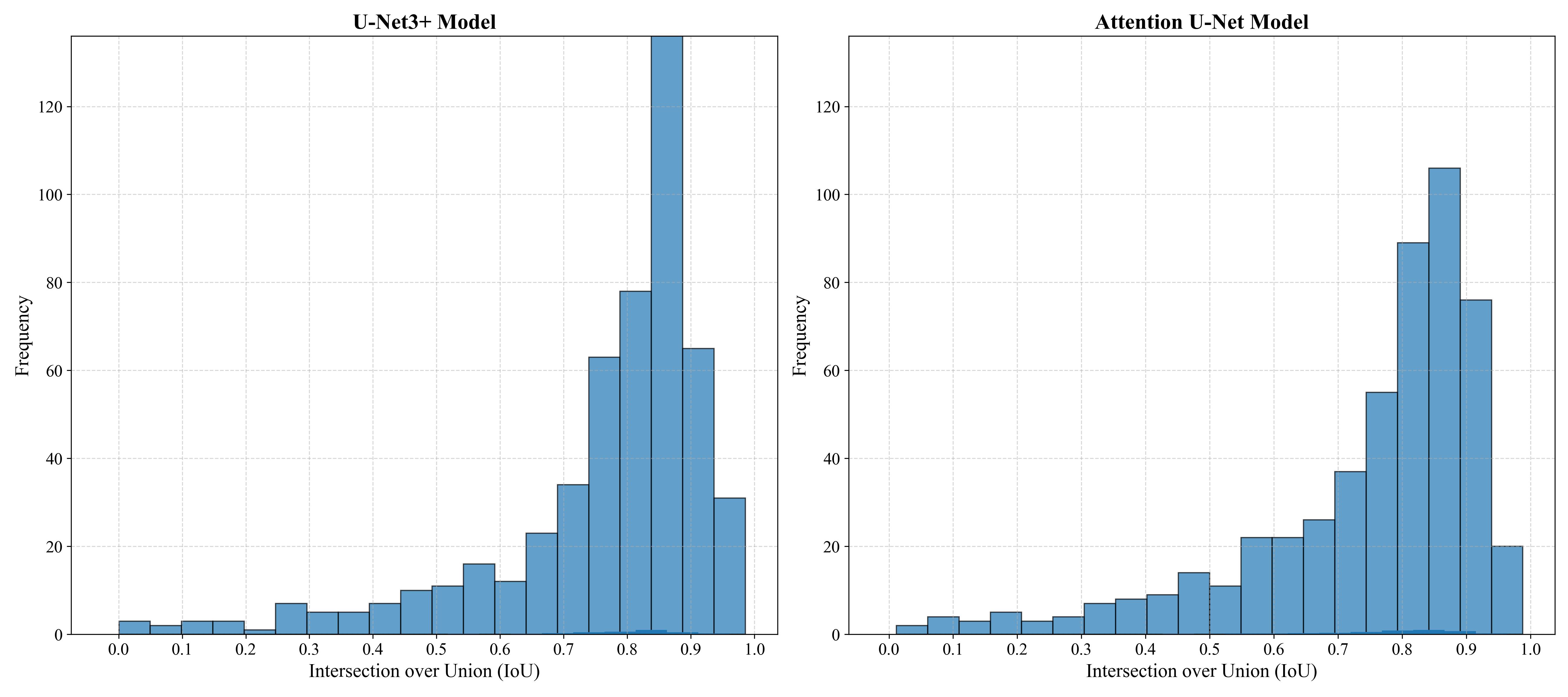}}
		\vspace{0.05cm}
		\caption{Distribution of IoU values achieved by different segmentation methods for building footprint detection}
		\label{fig_iouHist}
	\end{figure}
	
	Fig.\ref{fig_iouHist} displays the histogram of IoU value distributions for detected building footprints using the two models, Attention U-Net and U-Net3+. The x-axis represents the IoU values, which measure the overlap between the detected building footprints and the ground truth footprints. At the same time, the y-axis shows the number of buildings corresponding to each IoU bin. These histograms provide insights into the two segmentation models' performance in detecting building footprints. The results indicate that most IoU values are distributed in the higher range (between 0.8 and 0.9), demonstrating a significant overlap between the detected footprints and the ground truth.
	Notably, there are also differences in frequency distribution between the two models. The U-Net3+ model shows higher frequencies in specific bins, such as 0.75, 0.85, and 0.95, suggesting that more buildings have higher IoU values than the Attention U-Net model. Also, the Attention U-Net model has higher frequencies in other bins, like 0.60, 0.8, and 0.9. However, overall, the total number of footprints with higher IoU values is more significant for the U-Net3+ model than for the Attention U-Net model. This indicates that the U-Net3+ model better overlaps the detected footprints and the ground truth. The higher frequency of IoU values suggests that the U-Net3+ model outperforms the Attention U-Net model.

	\begin{table}[!t]
		\centering
		\caption{Height errors for 3D models generated using different statistical measures. The values represent the percentage frequency (\%) of buildings falling within each error range.}
		\label{table_HErrHist}
		\begin{adjustbox}{width=\textwidth}
			\begin{tabular}{c|c|cccccccccc}
				\multirow{2}{*}{Model} & \multirow{2}{*}{Z} & \multicolumn{10}{c}{Height error range (E) in meter} \\
				\cmidrule(lr){3-11}
				& & $E < 0.1$ & $0.1 < E < 0.2$ & $0.2 < E < 0.35$ & $0.35 < E < 0.5$ & $0.5 < E < 0.75$ & $0.75 < E < 1$ & $1 < E < 1.5$ & $1.5 < E < 2.5$ & $2.5 < E < 5$ & $5 < E$ \\
				\midrule
				\multirow{5}{*}{U-Net3+} & Median & 60.42 & 13.96 & 10.13 & 5.35 & 3.63 & 1.34 & 2.29 & 1.91 & 0.76 & 0.19 \\
				& Maximum & 90.49 & 2.72 & 1.55 & 0.97 & 1.36 & 0.58 & 0.58 & 0.58 & 0.58 & 0.58 \\
				& Mode & 79.03 & 1.75 & 2.52 & 1.94 & 2.14 & 0.78 & 2.72 & 2.33 & 5.05 & 1.75 \\
				& Range & 61.55 & 12.04 & 6.80 & 2.72 & 3.30 & 2.52 & 3.11 & 3.50 & 3.50 & 0.97 \\
				& 90\textsuperscript{th} Percentile & 87.96 & 6.41 & 1.55 & 0.58 & 1.55 & 0.39 & 0.19 & 0.19 & 0.58 & 0.58 \\
				\midrule
				\multirow{5}{*}{\shortstack{Attention\\ U-Net}} & Median & 60.00 & 17.48 & 10.29 & 3.88 & 3.11 & 0.58 & 2.14 & 0.97 & 1.36 & 0.19 \\
				& Maximum & 89.87 & 1.34 & 1.15 & 0.76 & 1.53 & 1.15 & 0.96 & 0.96 & 1.72 & 0.57 \\
				& Mode & 76.86 & 1.72 & 2.10 & 1.91 & 1.72 & 1.75 & 2.87 & 3.25 & 6.31 & 1.53 \\
				& Range & 60.61 & 12.05 & 4.59 & 3.25 & 4.40 & 2.10 & 2.49 & 4.78 & 4.40 & 1.34 \\
				& 90\textsuperscript{th} Percentile & 85.66 & 5.74 & 3.63 & 0.76 & 0.96 & 0.57 & 0.38 & 0.76 & 1.34 & 0.19 \\
			\end{tabular}
		\end{adjustbox}
	\end{table}

	Table \ref{table_HErrHist} presents the height errors for 3D models generated using different statistical measures. The table categorizes errors into different ranges and provides the percentage of buildings that fall within each error range for both the Attention U-Net and U-Net3+ models. The height errors were computed based on the differences between LOD1 models derived from footprints identified by the Attention U-Net and U-Net3+ segmentation models and the ground truth footprints (OSM). As shown in Table \ref{table_HErrHist}, most height error values are close to zero, indicating relatively accurate height estimates for most buildings using both models. Overall, both models show similar height estimation performance across different statistical measures. Among the statistical measures, the 90\textsuperscript{th} percentile criterion yields the highest frequency of buildings with lower errors, followed by the maximum measure.
	
	On the other hand, as presented in Table \ref{table_HeightError}, the median measure yielded higher accuracy in modeling according to RMSE value, while based on the MAE, and $R^{2}$ maximum measure is more accurate. Combining the results from Table \ref{table_HeightError} and Table \ref{table_HErrHist}, it can be concluded that using the maximum measure results in a 3D model with a higher frequency of buildings with lower height errors. In contrast, a few buildings exhibit very high errors, while these buildings have lower errors if the median measure is used. In other words, the median criterion is less affected by significant errors and misidentifications of buildings. Overall, the appropriate measure for 3D building modeling should be determined by the specific applications.
	
	\begin{figure}[!t]
		\centering
		{\includegraphics[width=0.65\linewidth]{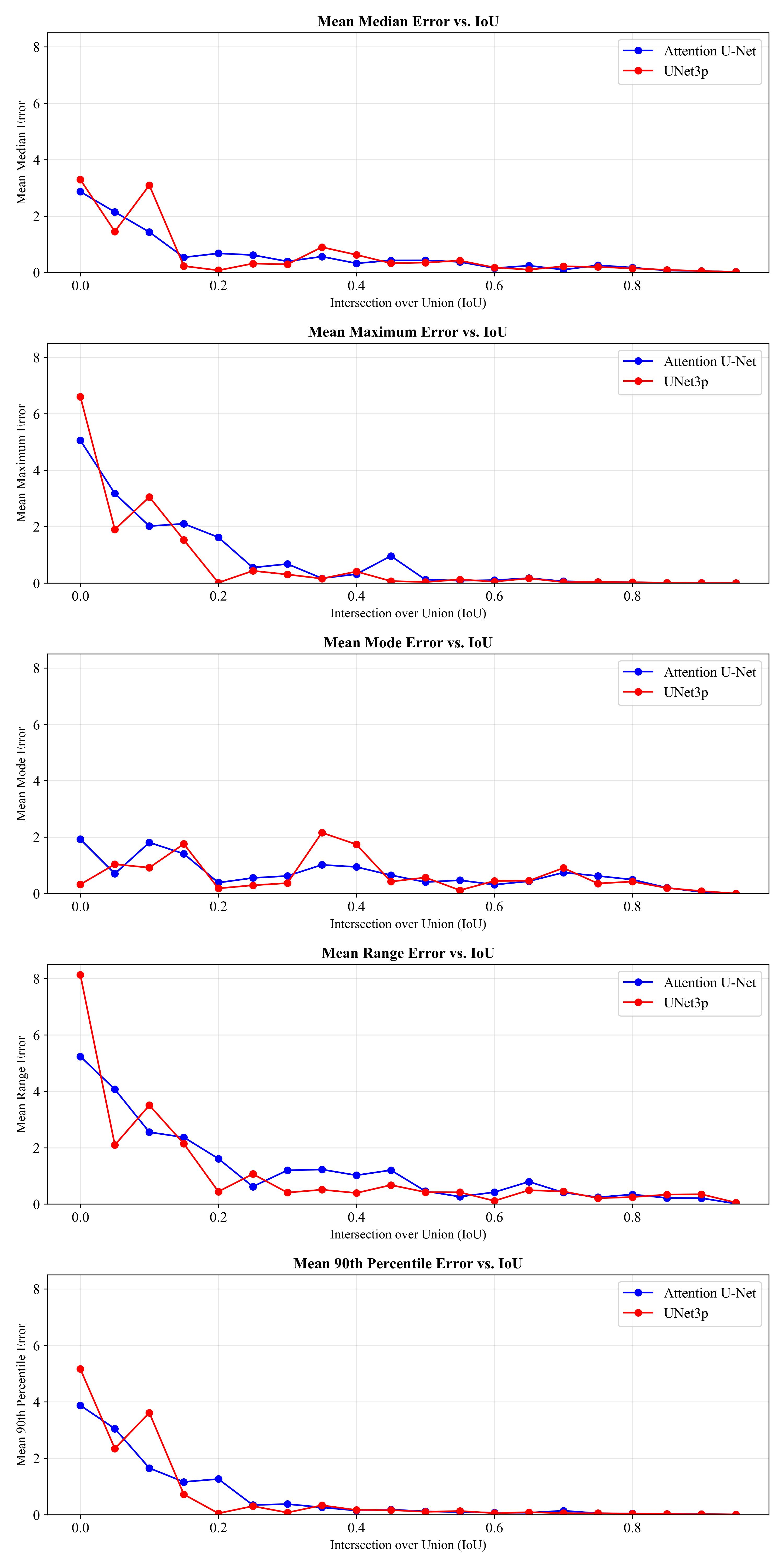}}
		\caption{Relationship between IoU (IoU Intervals of ±0.05) and MAE of heights for different 3D building models derived from building footprints recognized by U-Net3+ and Attention U-Net models and various height estimation measures.}
		\label{fig_iouHeight}
	\end{figure}
	
	Fig.\ref{fig_iouHeight} illustrates the relationship between the mean height error for five different height calculation measures and the IoU metric for footprints identified by the U-Net3+ and Attention U-Net models. In general, it is observed that higher errors are typically associated with lower IoU values. In other words, buildings with higher IoU values, which indicate a better match between the identified footprint and the ground truth, usually have lower height errors. This implies that buildings with higher IoU values are more accurately identified, leading to a more precise 3D model reconstruction and reduced height estimation errors. Additionally, the comparison between the two models reveals that buildings identified with the U-Net3+ model generally have lower height errors and higher IoU values than those identified using the Attention U-Net model. In other words, the U-Net3+ model performs better in detecting building footprints and estimating their heights, aligning with the results presented in Table \ref{table_HeightError}.
	
	On the other hand, in cases where the IoU is close to zero, the U-Net3+ model typically has higher height errors than the Attention U-Net model. This suggests a more realistic performance of the U-Net3+ model, as lower IoU values indicate less overlap between the identified and actual footprints, which should lead to more significant errors in the 3D model. In contrast, the Attention U-Net model results in a less accurate building reconstruction under these conditions.
	
	Among various statistical measures, the median, maximum, and 90\textsuperscript{th} percentile measures are more effective in modeling the relationship between height error and Intersection over Union (IoU). This is because, as the IoU increases, the height error tends to decrease, approaching zero. In the case of the 90\textsuperscript{th} percentile, the height error accuracy decreases with the influence of IoU, and when the IoU threshold exceeds 0.4, the height error approaches zero.
	
	\subsubsection{Impact of Employing Different Height Estimation Measures on 3D Building Reconstruction}\label{sub_3dHDiscuss}
	
	\begin{figure}[!t]
		\centering
		{\includegraphics[width=\linewidth]{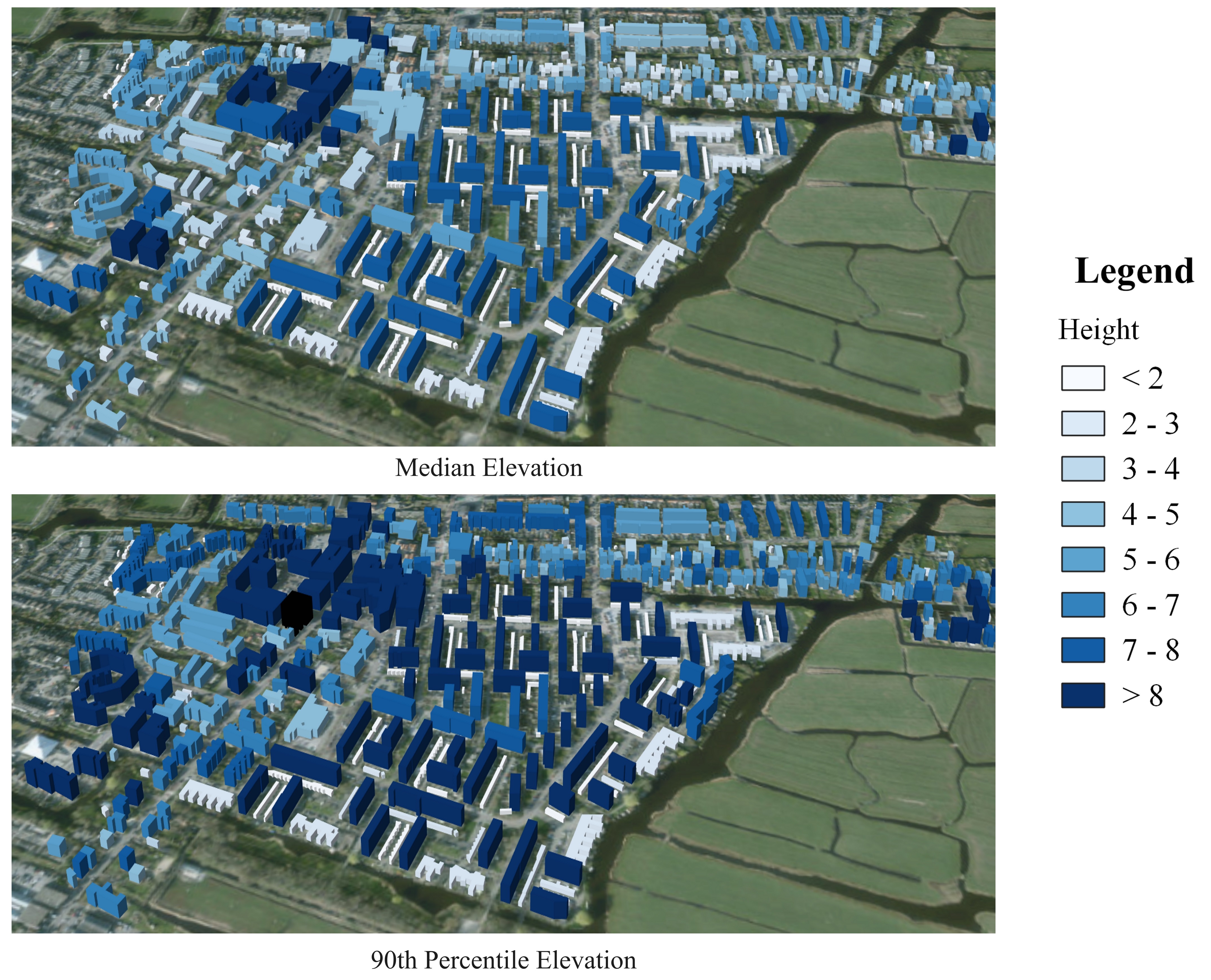}}
		\caption{The 3D building model at the LOD1 using various height estimation measures and footprints derived from U-Net3+}
		\label{fig_3DUnet3p}
	\end{figure}
	
	Different height estimation measures provide valid results depending on the building structure and application. For instance, the maximum height helps identify skyscrapers or detect obstacles for drones, but it can be skewed by outliers such as antennas. The range effectively identifies multi-story buildings but may produce misleading results in areas with uneven terrain. The mode performs well for flat-roofed structures but struggles with sloped or irregular rooftops. In contrast, the median and 90\textsuperscript{th} percentile offer more stable estimates in dense urban areas, as they are less sensitive to extreme values. Understanding these trade-offs ensures that the chosen measure aligns with the specific requirements of 3D modeling and urban analysis.
	
	Based on the results and evaluation of various height estimation metrics, the median and 90\textsuperscript{th} percentile were selected for further analysis. These measures were found to be less sensitive to data variability and outliers, showing better overall performance in height estimation. Therefore, they were chosen for the detailed 3D building reconstruction analysis.
	
	Fig.\ref{fig_3DUnet3p} illustrates the 3D building model at LOD1, where building heights are calculated using the 90\textsuperscript{th} percentile and median methods. In both scenarios, the footprints are derived from U-Net3+. Building heights are categorized and presented using the same symbology for both methods for better comparison. It can be observed that buildings calculated using the 90\textsuperscript{th} percentile measure tend to have higher heights than those calculated using the median measure. Statistical analysis also supports this observation,  indicating that both the minimum and maximum building heights are greater when calculated using the 90\textsuperscript{th} percentile measure compared to the median measure. Specifically, the minimum building height using the 90\textsuperscript{th} percentile is 1.39 m, and the maximum is 12.17 m, whereas the median measure results in a lower range, with a minimum height of 1.27 m and a maximum height of 9.86 m.
	
	The difference in building heights between the two calculation methods underscores the impact of the chosen statistical measure on building height estimation. The 90\textsuperscript{th} percentile measure tends to capture higher heights for each building, potentially reflecting the presence of outliers or extreme values in the height data. On the other hand, the median measure provides a measure of central tendency, which is less sensitive to outliers, leading to relatively lower building height estimates. However, it is important to note that while the 90\textsuperscript{th} percentile helps mitigate underestimations, it may still introduce slight biases by overemphasizing higher elevation points, particularly in buildings with complex rooftop structures. Similarly, while the median is robust against extreme values, it may smooth out meaningful height variations, potentially underrepresenting the actual vertical extent of some buildings. These biases, which will be discussed in detail in Section \ref{sub_DiscussLimit}, should be considered when selecting a height estimation method for specific urban applications.
	
	\subsubsection{Impact of Segmentation Accuracy on Urban Morphology Extraction}\label{sub_MorphologyDiscuss}
	
	This section examines the impact of building footprint segmentation accuracy on the accuracy of extracting urban morphological features. For this purpose, it explores the relationship between segmentation quality (measured by IoU) and the accuracy of area estimation. Additionally, it analyzes how the uncertainty in footprint extraction affects building height errors in 3D models, considering various height calculation measures and their impact on accurately estimating the external wall area of buildings.

\begin{figure}[!t]
		\centering
		{\includegraphics[width=\linewidth]{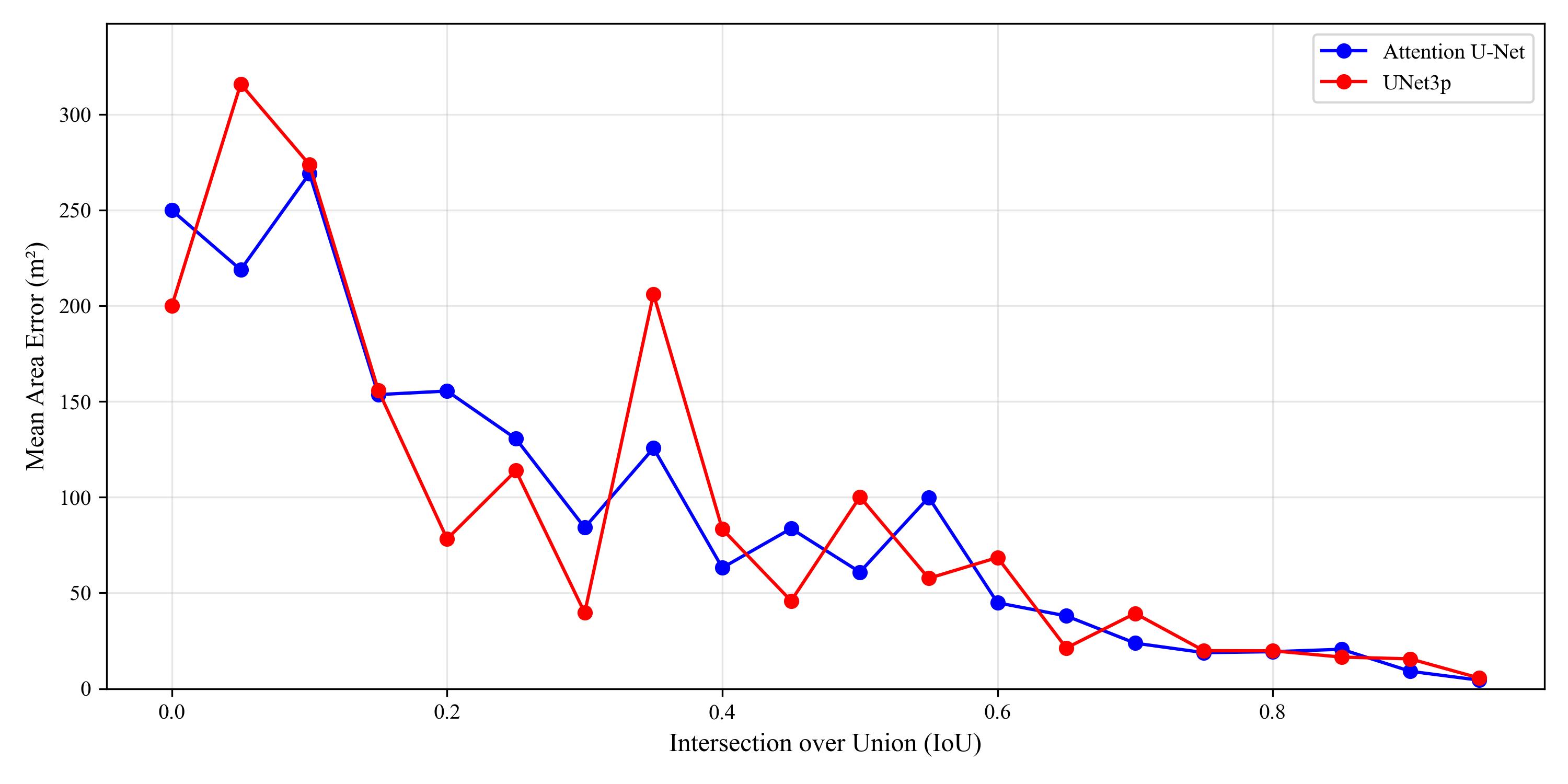}}
		\caption{Relationship between IoU and building footprint area}
		\label{fig_areaIoU}
	\end{figure}

Fig.\ref{fig_areaIoU} illustrates the relationship between area error and the IoU metric for the two models, Attention U-Net and U-Net3+. In this figure, each point represents the average area error for buildings whose IoU falls within a specific range. Generally, it is observed that as the IoU value increases, the area error decreases. Therefore, improving segmentation accuracy and achieving a closer match between detected and reference footprints lead to more precise area estimations.
Both models' area errors are relatively high for buildings with IoU values below 0.2, indicating a lack of accuracy in identifying these buildings. This suggests that poor segmentation quality results in more significant errors in building area estimation. Conversely, the area error decreases for buildings with IoU values above 0.5, indicating that higher segmentation accuracy leads to more accurate area estimates. Overall, both models exhibit similar behavior, showing that as segmentation quality improves (i.e., IoU increases), the accuracy of area estimation also improves.
	
\begin{figure}[!t]
		\centering
		{\includegraphics[width=\linewidth]{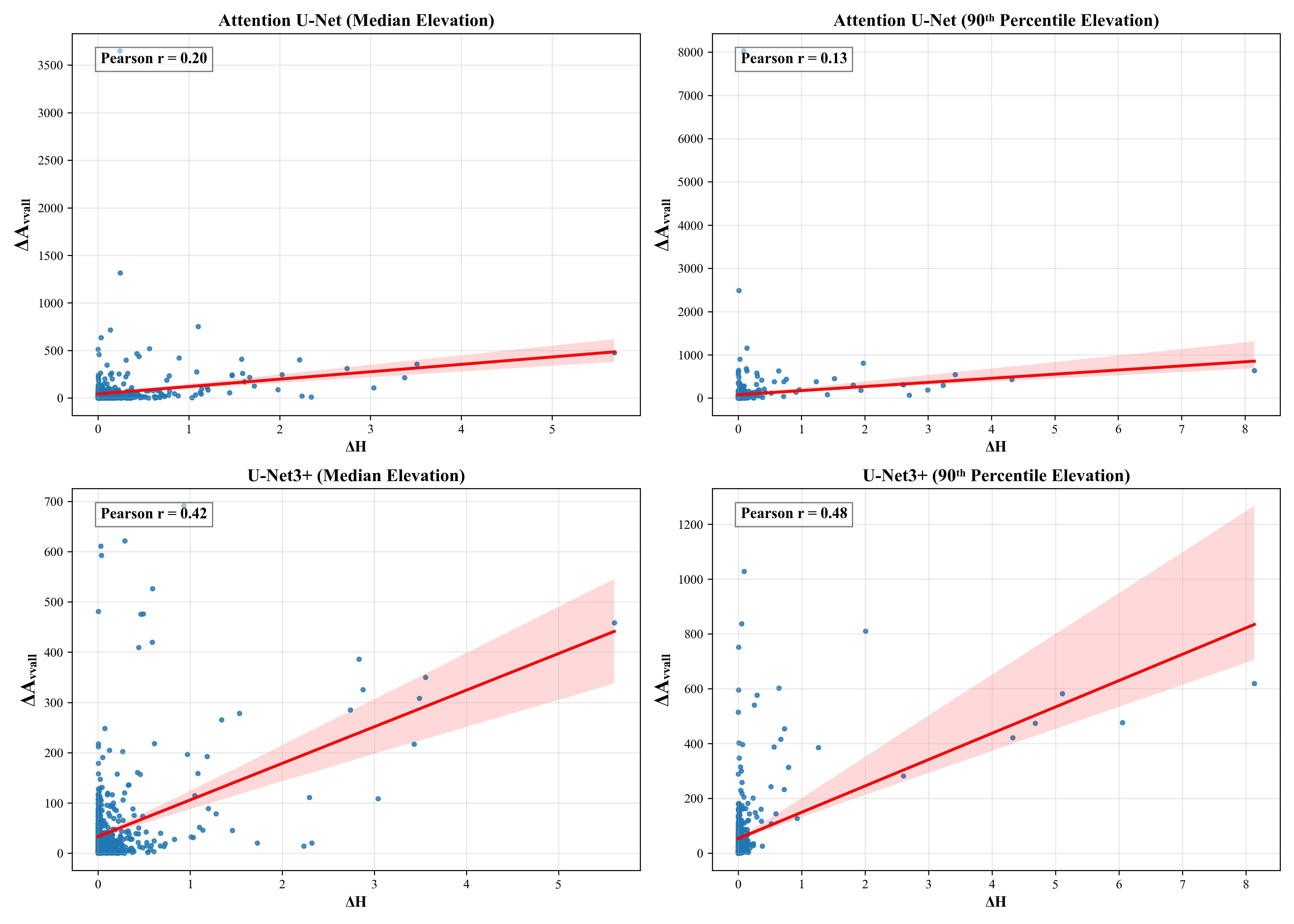}}
		\caption{Relationship between quality of 3d building model and the calculated areas of exterior walls}
		\label{fig_HWall}
	\end{figure}

	Fig.\ref{fig_HWall} illustrates the relationship between the height estimation error for each building in the 3D model due to uncertainty in footprint detection by the U-Net3+ and Attention U-Net models, and the accuracy of external wall area estimation, considering two height measures (median and 90\textsuperscript{th} percentile). This analysis used Pearson's correlation coefficient to assess the correlation between height errors and external wall area calculation errors. The results indicate a significant positive correlation between 3D model quality and external wall area errors. This means that an increase in height estimation error directly leads to a rise in external wall area estimation error.

The U-Net3+ model shows higher correlations in both height calculation methods (0.42 for the median and 0.48 for the 90\textsuperscript{th} percentile) compared to the Attention U-Net model (0.2 for the median and 0.13 for the 90\textsuperscript{th} percentile). In other words, the U-Net3+ model is more effective in capturing and modeling the variations and fluctuations in height errors due to uncertainties in footprint detection with corresponding wall area errors. This high correlation suggests that if there is an error in footprint detection, it is likely that the wall area error will also increase.

The primary reason is that the U-Net3+ model is more accurate in identifying and distinguishing building components, leading to more precise area estimations. This greater accuracy in building identification and differentiation ensures a more accurate correlation between height errors from footprint detection uncertainties and wall area errors. Another noteworthy point is the difference in the range of external wall area estimation errors between the models. The Attention U-Net model exhibits a broader range of errors, indicating lower accuracy in accurately identifying and distinguishing building components.
	
	\subsection{Comparative Analysis}\label{sub_CompareDiscuss}
In recent years, the integration of LiDAR data and deep learning techniques has been explored for building extraction and 3D reconstruction, with each approach presenting unique contributions and challenges. 

\citeauthor{RN343} employed a Fully Convolutional Network (FCN) for building detection from LiDAR data, then refined and regularized it into 2D primitives. The RANSAC algorithm was then applied to fit planar segments for rooftop shaping, enabling the formation of complete LOD1 3D building models. Their method effectively automated the reconstruction process. The reliance on RANSAC for plane fitting and boundary regularization achieved correctness and completeness above 72\%, but challenges like minor building omissions influenced the results \citep{RN343}. Similarly, \citeauthor{RN349} utilized RANSAC to detect roof planes from classified LiDAR data, but their method was limited by computational inefficiencies, as it processed point-based geometric primitives rather than well-defined planes. Their approach struggled with complex roof structures and required substantial computing power, making it less adaptable to datasets with varying densities \citep{RN349}. In contrast to the dependence on RANSAC, our work analyzed and applied different statistical measures for building height estimation, providing more flexibility and accuracy in building extrusion and morphological feature extraction.

\citeauthor{RN341} combined Dynamic Graph Convolutional Neural Networks (DGCNN) and Euclidean Clustering to segment buildings directly from raw LiDAR data. Their approach achieved segmentation accuracy ranging from 74.28\% to 93.55\% and IoU scores between 0.65 and 0.84, depending on dataset characteristics. However, datasets with closely spaced buildings reduced performance, with IoU dropping to 0.65–0.72 due to the challenges posed by tight gaps. This highlights the sensitivity of their method to spatial building arrangements \citep{RN341}. \citeauthor{RN351} also discussed similar challenges in segmenting adjacent buildings in complex urban environments, where closely situated structures often lead to difficulties in distinguishing individual entities \citep{RN351}. Similarly, when using the Attention U-Net model, our study faced difficulties with closely situated structures. These challenges were effectively addressed by the U-Net3+ model, which achieved higher accuracy. On the study area dataset, U-Net3+ delivered an IoU of 0.833 and pixel accuracy of 0.961; on the Miami-Dade dataset, an IoU of 0.853 and pixel accuracy of 0.970. Incorporating transfer learning played a crucial role in enhancing model performance across datasets with varying structural complexities more effectively.

Several studies have incorporated LiDAR data alongside aerial or satellite imagery to enhance building segmentation accuracy \citep{RN42, RN96, RN240, RN350, RN342}. For instance, \citeauthor{RN342} significantly improved building footprints and boundary delineation by integrating LiDAR data with RGB aerial imagery. Using RGB imagery alone, they achieved a maximum IoU of 0.7914 with Context Transfer UNet (CT-UNet). However, the addition of LiDAR data substantially boosted performance. The best result was obtained among various tested configurations with an ensemble model combining U-Net Baseline, U-Net (DenseNet201 backbone), and CT-UNet (EfficientNetV2S), achieving an IoU of 0.8964. Compared to our LiDAR-only approach, their results underscore the benefits of leveraging multimodal data and advanced techniques like TTA and ensemble modeling \citep{RN342}. However, the reliance on aerial imagery significantly increases data acquisition costs and processing complexity. Such methods often require expensive, high-quality sensors and significant computational resources for multi-source alignment and preprocessing. Our method demonstrates that high segmentation accuracy can still be achieved using only LiDAR data, eliminating dependencies on expensive high-resolution imagery while maintaining robust performance. This makes our approach particularly suitable for large-scale urban modeling in resource-constrained environments.

Many studies have also utilized satellite or aerial imagery for building extraction and 3D modeling, leveraging spectral and height information for feature extraction and classification \citep{RN36, RN35, baghfusion, RN34}. For instance,  \citeauthor{RN344} employed the Fused-FCN4S framework, which integrated multiple remote sensing data sources such as RGB, panchromatic (PAN), and nDSM images to extract building footprints with an IoU of approximately 68.1\%. The model effectively distinguished buildings from other above-ground objects, such as trees, while minimizing errors in building outlines. However, the study noted lower performance on the Istanbul dataset than the Munich dataset, which served as the primary training source. This discrepancy was attributed to variations in building density, architectural styles, and maximum building heights, highlighting challenges in generalizing models across diverse urban landscapes \citep{RN344}. Similarly, in our study, a model trained on the Miami-Dade dataset performed less effectively on the Landsmeer City test area, primarily due to differences in urban morphology and building characteristics. By employing transfer learning, we overcame these limitations, significantly improving model performance and ensuring better generalization across datasets with varying complexities. By leveraging pre-trained models, we reduced training time and computational costs, making our approach more scalable.

Overall, previous studies have primarily focused on improving the accuracy of semantic segmentation for generating 3D models. However, little attention has been given to evaluating how segmentation uncertainty, particularly in building footprint extraction, propagates into 3D modeling and affects the accuracy of derived morphological parameters.

Unlike prior approaches that integrate multiple data sources, such as satellite or aerial imagery alongside LiDAR, our study relies solely on LiDAR data. While conventional methods that fuse imagery and LiDAR have demonstrated high accuracy, they often depend on the availability and quality of multi-source data, require extensive data preprocessing, sensor calibration, and multi-source data alignment, which can introduce additional uncertainties and increase computational complexity. This makes our method more cost-effective while maintaining high accuracy, eliminating the need for additional datasets and the associated processing complexities. By leveraging LiDAR’s precise geometric information rather than spectral properties, our method avoids limitations associated with variations in lighting conditions and material reflectance, which often degrade performance in conventional image-based approaches.

To enhance processing efficiency and streamline 3D modeling, we employed state-of-the-art segmentation techniques combined with transfer learning. This approach significantly reduces computational costs associated with footprint extraction from LiDAR data while achieving higher accuracy than previous methods, ensuring more consistent segmentation accuracy across different datasets. By leveraging deep learning instead of traditional rule-based segmentation, our method minimizes manual intervention, reduces processing time, and ensures scalability for large-scale applications. This not only mitigates performance drops in previously unseen urban areas but also enhances the adaptability of our approach for large-scale applications. Table \ref{table_compare} presents a comparative analysis of our method against traditional approaches, highlighting its advantages in terms of efficiency, adaptability, and accuracy.

\begin{table}[!t]
	\centering
	\renewcommand{\arraystretch}{1.3}
	\caption{Comparison of our LiDAR-only method vs. traditional 3D building modelling approaches}
	\label{table_compare}
	\begin{adjustbox}{width=\textwidth}
		\begin{tabular}{>{\centering\arraybackslash}p{3.5cm}|p{6.8cm}|p{6.5cm}}
			\multicolumn{1}{c|}{Feature} & \multicolumn{1}{c|}{Traditional Methods} & \multicolumn{1}{c}{Our Method} \\
			\hline
			\toprule
			Data Requirements & Often require multiple data sources (e.g. aerial imagery), 
			increasing processing complexity & Relies solely on LiDAR data, making it more 
			cost-effective and efficient \\
			\hline
			Building Height Estimation & Rely on a limited number of metrics & Uses multiple 
			height estimation metrics for better accuracy \\
			\hline
			Segmentation Approach & LiDAR-only methods rely on plane fitting and rule-based 
			methods, often computationally intensive and less adaptable to varying datasets & 
			Deep learning (U-Net3+), improves segmentation accuracy \\
			\hline
			Generalization Across Datasets & Model performance varies across urban environments 
			due to architectural differences & Transfer learning enhances adaptability, ensuring 
			consistent segmentation accuracy across different cities \\
			\hline
			Morphological Feature Analysis & Primarily focuses on segmentation accuracy for 3D 
			modeling but lacks an in-depth evaluation of segmentation uncertainty impacts & 
			Assesses how segmentation uncertainty affects morphological parameters, enhancing 
			urban analysis reliability \\
			\hline
			Computational Efficiency & High computational cost due to point-based segmentation 
			and multiple data sources & More efficient processing using LiDAR-only deep learning 
			segmentation \\
		\end{tabular}
	\end{adjustbox}
\end{table}

	\subsection{Practical Implications and Challenges}\label{sub_SuggestDiscuss}
	\subsubsection{Potential Applications}\label{sub_DiscussApplic}
	
	The 3D building models at LOD1 developed in this study offer valuable applications across urban planning, environmental studies, and climate analysis. LOD1 models, known for their cost-effectiveness and simplicity, are widely adopted globally despite being the coarsest volumetric representation in the CityGML standard. Their accuracy enables diverse applications across multiple fields.
	
	These models are beneficial for optimizing solar panel placement on facades or rooftops, where precise 3D representations facilitate solar exposure calculations, supporting renewable energy initiatives and energy-efficient designs \citep{RN318, RN345, Hosseini_undated-qc}. They also play a crucial role in temperature downscaling and analyzing localized climatic effects, such as urban heat islands and wind speed variations, by modeling building morphology and airflow \citep{chajaei2024machine}.
	
	Further applications include traffic noise analysis, enabling the assessment of quality-of-life impacts and noise barrier placement, and shadow analysis in areas with flat roofs \citep{RN317, RN332, RN346, RN347}. LOD1 models are instrumental in real estate mass valuation, providing insights into urban dynamics and property trends. Moreover, these models ensure accurate elevation data in flood risk assessments, critical for urban resilience planning \citep{RN300}.
	
	By reducing segmentation and height estimation uncertainties, LOD1 models expand their applicability across disciplines, empowering urban planners and environmental scientists to develop sustainable solutions and effectively address climate challenges.
	
	Extracting morphological features adds another utility layer to these models, particularly in applications requiring detailed structural data. These features are invaluable for energy consumption modeling, providing information about heating, cooling, and insulation needs. Precise measurements of wall areas and heights are critical for evaluating energy efficiency retrofits, such as installing insulation materials or green facades, to improve thermal performance and reduce urban carbon footprints.
	
	In land use and land cover (LULC) classification, building morphology enables more precise differentiation of zones, such as residential, commercial, or industrial areas, improving urban zoning and development planning. Similarly, these features contribute to local climate zone (LCZ) classification, which focuses on the thermal characteristics of urban areas. Parameters like building height and footprint area (building ratio) are essential for understanding urban heat distribution and designing interventions to mitigate localized climate impacts, such as optimizing street canyons or implementing green infrastructure \citep{RN348}. Precise calculation of building areas and exterior wall surfaces also supports population density analysis and zoning efficiency studies, assisting urban planners in allocating land effectively. These features further aid biodiversity enhancement efforts by identifying opportunities for vertical greening systems and other sustainable urban designs.
	
	Moreover, by quantifying building areas, the models can be used for disaster risk assessments, such as evaluating structural vulnerabilities in flood-prone or earthquake-affected regions. These applications highlight the transformative potential of integrating morphological data into decision-making processes, promoting brighter, greener, and more resilient urban environments.
	
	\subsubsection{Limitations}\label{sub_DiscussLimit}
	
	While LOD1 models effectively serve many urban applications, their simplicity limits the inclusion of finer architectural details like roof types and facade characteristics, better captured in higher levels of detail, such as LOD2 or LOD3. Incorporating these higher levels could provide more comprehensive insights into building morphology and further expand the scope of their applications.
	
	As discussed, each statistical height estimation method is suited to specific building structure and application, but each also introduces potential inaccuracies. For example, the maximum height measure may capture unnecessary elements like antennas or chimneys, leading to overestimated heights. The range measure, influenced by both maximum and minimum points, can be misleading in hilly areas or when ground points are misclassified, requiring careful ground height definition. The mode is less effective for sloped or irregular roofs, as it may only reflect the most common height without capturing variations. In contrast, the median provides a robust central estimate that minimizes the influence of outliers but may overlook extreme height variations. Similarly, the 90\textsuperscript{th} percentile offers a more stable height estimate by excluding extreme values, though the accuracy can vary depending on the chosen percentile (e.g., 90\% vs. 95\%). 
	
	These inaccuracies can directly impact real-world applications. Overestimated building heights may result in incorrect floor area ratios, affecting zoning and land-use planning decisions. Underestimations could impact energy modeling accuracy, potentially leading to incorrect heating and cooling demand predictions. For example, in energy modeling, misestimated heights can alter calculations of solar exposure and shading effects, compromising the efficiency of passive solar designs. Similarly, height estimation errors in disaster management applications could impact risk assessment and evacuation planning, particularly in high-rise or flood-prone areas.
	
	The choice of height estimation method directly impacts not only the accuracy of 3D building modeling but also the accuracy of extracted morphological parameters. Using inaccurate height measures can lead to overestimation or underestimation of these features, affecting their reliability for different urban scenarios. Therefore, selecting an appropriate height metric based on the application context is crucial for ensuring reliable urban analysis.
	
	Additionally, this study examined a limited number of morphological parameters, including building height, footprint area, and exterior wall area. While these parameters are fundamental, expanding the scope to include other morphological features—such as volume-to-surface ratio, building orientation, floor area ratio, and building compactness—could provide a more comprehensive understanding of urban structures. Furthermore, the impact of uncertainty in LOD1 modeling on the accuracy of these additional morphological features should be investigated.
	
	Moreover, inconsistencies between OSM labels and LiDAR data introduce uncertainties in 3D building modeling. Differences in data collection times, along with potential inaccuracies in OSM labels due to their crowdsourced nature, can lead to missing, misaligned, or erroneous footprints, impacting the accuracy of reconstructed models. We tried to identify and mitigate these inconsistencies by selecting more accurate labels whenever possible. Additionally, using a DSM instead of point clouds introduces interpolation effects, sometimes resulting in jagged building edges. While post-processing techniques were applied to minimize these distortions, some uncertainties remain. Future work could explore integrating more precise cadastral data and refining interpolation methods to improve footprint accuracy further.
	
While LiDAR data provides high-precision height measurements, its reliance as the sole data source limits applicability in areas with sparse coverage. Alternative data sources, such as photogrammetric DSMs from satellite and aerial imagery, stereo imagery, or radar-based height models, could extend the approach to regions without LiDAR availability. However, these methods introduce higher noise levels compared to LiDAR and often require more computationally intensive processing to achieve comparable accuracy. Additionally, challenges such as occlusions, lower vertical resolution, and surface smoothing effects can impact the reliability of height estimation in complex urban environments.
	
	Furthermore, the generalizability of our model may be influenced by the specific characteristics of the study area. Differences in building structures across regions could introduce uncertainties in footprint extraction or height estimation. For instance, certain building types may be better represented using alternative height metrics based on their architectural characteristics. 
	
	This highlights the need for localized calibration of height estimation methods to account for regional variations in urban morphology. For example, in dense urban areas with high-rise buildings, methods that emphasize upper percentiles (e.g., 90\textsuperscript{th} percentile) may be more reliable, whereas in suburban or low-rise areas, median-based measures may better capture representative heights.
	
	Similarly, in regions dominated by flat-roofed buildings, single-value estimates such as the median or mode can provide reliable results. However, for regions with a prevalence of pitched or sloped roofs, these methods may fail to capture height variations accurately, and alternative approaches, such as percentile-based estimation, might be more appropriate.
	
	Future research could evaluate model performance in diverse urban settings by applying and validating different height estimation techniques. Testing across cities with varying architectural styles and terrain conditions would help refine the methodology and improve adaptability to a wider range of urban scenarios.

	\subsection{Future Directions}\label{sub_DiscussFuture}
	
	Future research could address these limitations by exploring higher levels of detail (LOD2 or LOD3) to incorporate additional architectural features, such as roof shapes and facade characteristics. This would enable more precise applications, such as detailed energy consumption simulations or advanced urban design studies. Expanding the scope of morphological parameters in future studies could further enhance the utility of the models in diverse fields. 
	
	Moreover, the study did not explicitly analyze how uncertainties in building footprint extraction or height estimation propagate into practical applications, such as temperature modeling, land-use classification, or wind flow analysis. Understanding and quantifying these uncertainties can be highly beneficial for improving the reliability and applicability of the generated models. A comprehensive uncertainty analysis could evaluate the impact of these inaccuracies on various applications. For example, exploring how uncertainties in building area measurements affect temperature modeling would provide valuable insights into its effects on urban climate predictions. Similarly, analyzing the implications of these uncertainties on LCZ classification, urban energy modeling, or solar panel deployment would help refine current methodologies. These efforts would contribute to developing more accurate and robust 3D building models for various urban and environmental applications.
	
	Another proposal for future research is investigating alternative statistical measures for estimating building heights. While this study employed several metrics, including the 90\textsuperscript{th} percentile, further exploration is needed to evaluate the suitability of other height percentiles or statistical parameters. Adjusting the height percentile based on the morphology and typology of buildings in specific regions could yield more accurate height estimations tailored to local urban contexts. 
	
	Additionally, integrating complementary data sources, such as aerial images or multispectral satellite data, could enhance model accuracy, particularly for higher levels of detail, where finer architectural elements such as roof geometry and facade structures become more relevant. Such an approach would provide greater flexibility and precision in height modeling, ultimately enhancing the overall reliability of the generated 3D models.

	\section{Conclusion}\label{sect_conclusion}
	This study demonstrated the effectiveness of using LiDAR data and deep learning models for building footprint extraction and 3D reconstruction. It highlighted the impact of segmentation accuracy on 3D modeling and morphology extraction. Among the four tested deep semantic segmentation models—U-Net, Attention U-Net, U-Net3+, and DeepLabV3+—U-Net3+ and Attention U-Net delivered superior performance in extracting building footprints. The application of transfer learning significantly enhanced model accuracy, with U-Net3+ achieving the best results in detecting building footprints. Attention U-Net encountered challenges in distinguishing adjacent buildings, leading to the potential merging of footprints and increased height estimation errors.
	
	The analysis revealed that uncertainties in building footprint detection significantly influenced building height estimation and 3D model quality. Building heights were derived from LiDAR point clouds using different statistical measures. The results revealed that the median and 90\textsuperscript{th} percentile measures provided the most reliable height estimations, while the mode metric underperformed. Buildings with higher IoU values were shown to have lower height errors, emphasizing the direct correlation between segmentation accuracy and reliable height estimation. These uncertainties also influenced morphological parameters, such as building areas and the total surface area of external walls, where inaccuracies in footprint detection and height estimation could lead to varying degrees of error in parameter calculations.
	
	This study also demonstrated that the choice of statistical measures for height estimation, such as the median or 90\textsuperscript{th} percentile, directly impacts the accuracy of extracted morphological parameters like the external wall area. Overall, segmentation accuracy significantly influences the quality of 3D building models and subsequent morphological analyses.
	
	In summary, this study explored the feasibility of extracting buildings from LiDAR data, reconstructing them in 3D at LOD1, and extracting building morphological features. The results confirm that integrating LiDAR data with deep learning models is a powerful approach for building footprint extraction and 3D reconstruction. The findings showed that U-Net3+ outperformed other models in footprint detection, leading to more accurate 3D models and morphological extractions. The median and 90\textsuperscript{th} percentile measures were the most reliable for height estimation, minimizing errors in model reconstruction. Furthermore, the accuracy of footprint detection played a critical role in determining building area and external wall surface estimations. By examining segmentation uncertainties and their impact on modeling processes, this study provides valuable insights for improving urban analysis methodologies. The findings can be applied to various fields, such as urban planning, designing communication networks, and environmental management, thereby contributing to more informed decision-making processes. Additionally, analyzing building morphology using these data and models can provide valuable insights for a better understanding of the structure and shape of cities, aiding in more precise and efficient urban planning.
	
	\section*{Acknowledgement}
	We would like to express our gratitude to \href{https://www.openstreetmap.org/}{OpenStreetMap} for providing essential geospatial data and \href{https://ahn.nl/}{AHN} for supplying the Amsterdam LiDAR data.
	
	\section*{Data and Code Availability}
	The code used in this study is openly available in the following GitHub repository: \href{https://github.com/FatemehCh97/LiDAR-3D-Building-Modeling}{GitHub Repository}.
	
	Both LiDAR datasets used in this study are also publicly available. The Netherlands dataset (AHN3) can be accessed at \href{https://www.ahn.nl/dataroom}{AHN3 Data Portal}, while the Miami-Dade dataset is available in DEM format at \href{https://www.fisheries.noaa.gov/inport/item/59009}{NOAA Fisheries}.
	
	\section*{Author contributions: CRediT}
	\textbf{Fatemeh Chajaei:} Writing – original draft, Visualization, Software, Formal analysis, Data curation, Validation. \textbf{Hossein Bagheri:} Writing – review and editing, Software, Validation, Supervision, Methodology, Investigation, Conceptualization.
	
	\clearpage
	\bibliographystyle{model1-num-names}
	\bibliography{ReferenceLibrary.bib}

\begin{thebibliography}{99}
\expandafter\ifx\csname natexlab\endcsname\relax\def\natexlab#1{#1}\fi
\providecommand{\bibinfo}[2]{#2}
\ifx\xfnm\relax \def\xfnm[#1]{\unskip,\space#1}\fi
\bibitem[{Ketzler et~al.(2020)Ketzler, Naserentin, Latino, Zangelidis,
  Thuvander, and Logg}]{RN55}
\bibinfo{author}{B.~Ketzler}, \bibinfo{author}{V.~Naserentin},
  \bibinfo{author}{F.~Latino}, \bibinfo{author}{C.~Zangelidis},
  \bibinfo{author}{L.~Thuvander}, \bibinfo{author}{A.~Logg},
\newblock \bibinfo{title}{Digital twins for cities: A state of the art review},
\newblock \bibinfo{journal}{Built Environment} \bibinfo{volume}{46}
  (\bibinfo{year}{2020}) \bibinfo{pages}{547--573}.
\bibitem[{Singh et~al.(2013)Singh, Jain, and Mandla}]{RN27}
\bibinfo{author}{S.~P. Singh}, \bibinfo{author}{K.~Jain},
  \bibinfo{author}{V.~R. Mandla},
\newblock \bibinfo{title}{Virtual {3D} city modeling: Techniques and
  applications},
\newblock \bibinfo{journal}{Int. Arch. Photogramm. Remote Sens. Spatial Inf.
  Sci.} \bibinfo{volume}{XL-2/W2} (\bibinfo{year}{2013})
  \bibinfo{pages}{73--91}. \bibinfo{note}{ISPRS-Archives}.
\bibitem[{Biljecki et~al.(2015)Biljecki, Stoter, Ledoux, Zlatanova, and
  Çöltekin}]{RN47}
\bibinfo{author}{F.~Biljecki}, \bibinfo{author}{J.~Stoter},
  \bibinfo{author}{H.~Ledoux}, \bibinfo{author}{S.~Zlatanova},
  \bibinfo{author}{A.~Çöltekin},
\newblock \bibinfo{title}{Applications of {3D} city models: State of the art
  review},
\newblock \bibinfo{journal}{ISPRS International Journal of Geo-Information}
  \bibinfo{volume}{4} (\bibinfo{year}{2015}) \bibinfo{pages}{2842--2889}.
\bibitem[{Ross(2011)}]{RN303}
\bibinfo{author}{L.~Ross},
\newblock \bibinfo{title}{Virtual {3D} city models in urban land
  management-technologies and applications}  (\bibinfo{year}{2011}).
\bibitem[{Czyńska and Rubinowicz(2014)}]{RN309}
\bibinfo{author}{K.~Czyńska}, \bibinfo{author}{P.~Rubinowicz},
\newblock \bibinfo{title}{Application of {3D} virtual city models in urban
  analyses of tall buildings : today practice and future challenges},
\newblock \bibinfo{journal}{Architecturae et Artibus} \bibinfo{volume}{Vol. 6,
  no. 1} (\bibinfo{year}{2014}) \bibinfo{pages}{9--13}.
\bibitem[{Kaňuk et~al.(2015)Kaňuk, Gallay, and Hofierka}]{RN308}
\bibinfo{author}{J.~Kaňuk}, \bibinfo{author}{M.~Gallay},
  \bibinfo{author}{J.~Hofierka},
\newblock \bibinfo{title}{Generating time series of virtual {3D} city models
  using a retrospective approach},
\newblock \bibinfo{journal}{Landscape and Urban Planning} \bibinfo{volume}{139}
  (\bibinfo{year}{2015}) \bibinfo{pages}{40--53}.
\bibitem[{Willenborg et~al.(2018)Willenborg, Sindram, and Kolbe}]{RN302}
\bibinfo{author}{B.~Willenborg}, \bibinfo{author}{M.~Sindram},
  \bibinfo{author}{T.~H. Kolbe},
\newblock \bibinfo{title}{Applications of {3D} city models for a better
  understanding of the built environment},
\newblock \bibinfo{journal}{Trends in spatial analysis and modelling:
  decision-support and planning strategies}  (\bibinfo{year}{2018})
  \bibinfo{pages}{167--191}.
\bibitem[{Barrile et~al.(2023)Barrile, Genovese, and Favasuli}]{RN318}
\bibinfo{author}{V.~Barrile}, \bibinfo{author}{E.~Genovese},
  \bibinfo{author}{F.~Favasuli},
\newblock \bibinfo{title}{Development and application of an integrated
  {BIM-GIS} system for the energy management of buildings},
\newblock \bibinfo{journal}{WSEAS TRANSACTIONS ON POWER SYSTEMS}
  (\bibinfo{year}{2023}).
\bibitem[{Sulistyah and Hong(2019)}]{RN301}
\bibinfo{author}{U.~D. Sulistyah}, \bibinfo{author}{J.-H. Hong},
\newblock \bibinfo{title}{The use of {3D} building data for disaster
  management: A {3D} {SDI} perspective},
\newblock \bibinfo{journal}{The International Archives of the Photogrammetry,
  Remote Sensing and Spatial Information Sciences} \bibinfo{volume}{XLII-3/W8}
  (\bibinfo{year}{2019}) \bibinfo{pages}{395--402}.
\bibitem[{Hong and Tsai(2020)}]{RN300}
\bibinfo{author}{J.-H. Hong}, \bibinfo{author}{C.-Y. Tsai},
\newblock \bibinfo{title}{Using {3D} {WebGIS} to support the disaster
  simulation, management and analysis – examples of tsunami and flood},
\newblock \bibinfo{journal}{The International Archives of the Photogrammetry,
  Remote Sensing and Spatial Information Sciences}
  \bibinfo{volume}{XLIV-3/W1-2020} (\bibinfo{year}{2020})
  \bibinfo{pages}{43--50}.
\bibitem[{Musliman et~al.(2006)Musliman, Rahman, and Coors}]{RN304}
\bibinfo{author}{I.~A. Musliman}, \bibinfo{author}{A.~A. Rahman},
  \bibinfo{author}{V.~Coors}, \bibinfo{title}{{3D} Navigation for {3D-GIS} ---
  Initial Requirements}, \bibinfo{publisher}{Springer Berlin Heidelberg},
  \bibinfo{address}{Berlin, Heidelberg}, pp. \bibinfo{pages}{259--268}.
\bibitem[{C{\"o}ltekin et~al.(2015)C{\"o}ltekin, Lokka, and Bo{\'e}r}]{RN305}
\bibinfo{author}{A.~C{\"o}ltekin}, \bibinfo{author}{I.-E. Lokka},
  \bibinfo{author}{A.~Bo{\'e}r},
\newblock \bibinfo{title}{The utilization of publicly available map types by
  non-experts—a choice experiment},
\newblock in: \bibinfo{booktitle}{Proceedings of the 27th International
  Cartographic Conference (ICC2015), Rio de Janeiro, Brazil}, pp.
  \bibinfo{pages}{23--28}.
\bibitem[{Nedkov(2012)}]{RN306}
\bibinfo{author}{S.~Nedkov},
\newblock \bibinfo{title}{Knowledge-based optimisation of {3D} city models for
  car navigation devices},
\newblock \bibinfo{journal}{Master’s Thesis}  (\bibinfo{year}{2012}).
\bibitem[{Chen and Gao(2019)}]{RN307}
\bibinfo{author}{D.~Chen}, \bibinfo{author}{G.~X. Gao},
\newblock \bibinfo{title}{Probabilistic graphical fusion of {LiDAR}, {GPS}, and
  {3D} building maps for urban {UAV} navigation},
\newblock \bibinfo{journal}{Navigation} \bibinfo{volume}{66}
  (\bibinfo{year}{2019}) \bibinfo{pages}{151--168}.
\bibitem[{Carrión et~al.(2010)Carrión, Lorenz, and Kolbe}]{RN311}
\bibinfo{author}{D.~Carrión}, \bibinfo{author}{A.~Lorenz},
  \bibinfo{author}{T.~H. Kolbe},
\newblock \bibinfo{title}{Estimation of the energetic rehabilitation state of
  buildings for the city of {Berlin} using a {3D} city model represented in
  {CityGML}},
\newblock in: \bibinfo{booktitle}{ISPRS Conference: International Conference on
  3D Geoinformation. XXXVIII-4}, volume \bibinfo{volume}{XXXVIII-4/W15},
  \bibinfo{publisher}{ISPRS}, \bibinfo{year}{2010}.
\bibitem[{Bahu et~al.(2013)Bahu, Koch, Kremers, and Murshed}]{RN313}
\bibinfo{author}{J.-M. Bahu}, \bibinfo{author}{A.~Koch},
  \bibinfo{author}{E.~Kremers}, \bibinfo{author}{S.~M. Murshed},
\newblock \bibinfo{title}{Towards a {3D} spatial urban energy modelling
  approach},
\newblock \bibinfo{journal}{ISPRS Annals of the Photogrammetry, Remote Sensing
  and Spatial Information Sciences} \bibinfo{volume}{II-2/W1}
  (\bibinfo{year}{2013}) \bibinfo{pages}{33--41}.
\bibitem[{Kaden and Kolbe(2014)}]{RN310}
\bibinfo{author}{R.~Kaden}, \bibinfo{author}{T.~H. Kolbe},
\newblock \bibinfo{title}{Simulation-based total energy demand estimation of
  buildings using semantic {3D} city models},
\newblock \bibinfo{journal}{International Journal of 3-D Information Modeling
  (IJ3DIM)} \bibinfo{volume}{3} (\bibinfo{year}{2014}) \bibinfo{pages}{35--53}.
\bibitem[{Wang et~al.(2021)Wang, Wei, Du, Zhuang, Li, Shi, Jin, and
  Zhou}]{RN312}
\bibinfo{author}{C.~Wang}, \bibinfo{author}{S.~Wei}, \bibinfo{author}{S.~Du},
  \bibinfo{author}{D.~Zhuang}, \bibinfo{author}{Y.~Li},
  \bibinfo{author}{X.~Shi}, \bibinfo{author}{X.~Jin},
  \bibinfo{author}{X.~Zhou},
\newblock \bibinfo{title}{A systematic method to develop three dimensional
  geometry models of buildings for urban building energy modeling},
\newblock \bibinfo{journal}{Sustainable Cities and Society}
  \bibinfo{volume}{71} (\bibinfo{year}{2021}) \bibinfo{pages}{102998}.
\bibitem[{Johari et~al.(2020)Johari, Peronato, Sadeghian, Zhao, and
  Widén}]{RN314}
\bibinfo{author}{F.~Johari}, \bibinfo{author}{G.~Peronato},
  \bibinfo{author}{P.~Sadeghian}, \bibinfo{author}{X.~Zhao},
  \bibinfo{author}{J.~Widén},
\newblock \bibinfo{title}{Urban building energy modeling: State of the art and
  future prospects},
\newblock \bibinfo{journal}{Renewable and Sustainable Energy Reviews}
  \bibinfo{volume}{128} (\bibinfo{year}{2020}) \bibinfo{pages}{109902}.
\bibitem[{Kurakula et~al.(2007)Kurakula, Skidmore, Kluijver, Stoter,
  Dabrowska~Zielinska, and Kuffer}]{RN317}
\bibinfo{author}{V.~Kurakula}, \bibinfo{author}{A.~Skidmore},
  \bibinfo{author}{H.~Kluijver}, \bibinfo{author}{J.~Stoter},
  \bibinfo{author}{K.~Dabrowska~Zielinska}, \bibinfo{author}{M.~Kuffer},
\newblock \bibinfo{title}{A {GIS} based approach for {3D} noise modelling using
  {3D} city models},
\newblock \bibinfo{organization}{ITC Enschede, The Netherlands}.
\bibitem[{Chajaei and Bagheri(2024)}]{chajaei2024machine}
\bibinfo{author}{F.~Chajaei}, \bibinfo{author}{H.~Bagheri},
\newblock \bibinfo{title}{Machine learning framework for high-resolution air
  temperature downscaling using {LiDAR}-derived urban morphological features},
\newblock \bibinfo{journal}{Urban Climate} \bibinfo{volume}{57}
  (\bibinfo{year}{2024}) \bibinfo{pages}{102102}.
\bibitem[{Verbree et~al.(1999)Verbree, Maren, Germs, Jansen, and Kraak}]{RN315}
\bibinfo{author}{E.~Verbree}, \bibinfo{author}{G.~V. Maren},
  \bibinfo{author}{R.~Germs}, \bibinfo{author}{F.~Jansen},
  \bibinfo{author}{M.-J. Kraak},
\newblock \bibinfo{title}{Interaction in virtual world views-linking {3D} {GIS}
  with {VR}},
\newblock \bibinfo{journal}{International Journal of Geographical Information
  Science} \bibinfo{volume}{13} (\bibinfo{year}{1999})
  \bibinfo{pages}{385--396}.
\bibitem[{{Marina, O.} et~al.(2012){Marina, O.}, {Masala, E.}, {Pensa, S.}, and
  {Stavric, M.}}]{RN316}
\bibinfo{author}{{Marina, O.}}, \bibinfo{author}{{Masala, E.}},
  \bibinfo{author}{{Pensa, S.}}, \bibinfo{author}{{Stavric, M.}},
  \bibinfo{title}{Interactive model of urban development in residential areas
  in skopje}, \bibinfo{year}{2012}.
\bibitem[{Somanath et~al.(2023)Somanath, Naserentin, Eleftheriou, Sjölie,
  Wästberg, and Logg}]{RN319}
\bibinfo{author}{S.~Somanath}, \bibinfo{author}{V.~Naserentin},
  \bibinfo{author}{O.~Eleftheriou}, \bibinfo{author}{D.~Sjölie},
  \bibinfo{author}{B.~S. Wästberg}, \bibinfo{author}{A.~Logg},
  \bibinfo{title}{On procedural urban digital twin generation and visualization
  of large scale data}, \bibinfo{year}{2023}.
\bibitem[{Bagheri et~al.(2018)Bagheri, Schmitt, d’Angelo, and
  Zhu}]{BAGHERI2018389}
\bibinfo{author}{H.~Bagheri}, \bibinfo{author}{M.~Schmitt},
  \bibinfo{author}{P.~d’Angelo}, \bibinfo{author}{X.~X. Zhu},
\newblock \bibinfo{title}{A framework for {SAR}-optical stereogrammetry over
  urban areas},
\newblock \bibinfo{journal}{ISPRS Journal of Photogrammetry and Remote Sensing}
  \bibinfo{volume}{146} (\bibinfo{year}{2018}) \bibinfo{pages}{389--408}.
\bibitem[{Sirmacek et~al.(2012)Sirmacek, Taubenbock, Reinartz, and
  Ehlers}]{RN42}
\bibinfo{author}{B.~Sirmacek}, \bibinfo{author}{H.~Taubenbock},
  \bibinfo{author}{P.~Reinartz}, \bibinfo{author}{M.~Ehlers},
\newblock \bibinfo{title}{Performance evaluation for {3D} city model generation
  of six different {DSMs} from air- and spaceborne sensors},
\newblock \bibinfo{journal}{IEEE Journal of Selected Topics in Applied Earth
  Observations and Remote Sensing} \bibinfo{volume}{5} (\bibinfo{year}{2012})
  \bibinfo{pages}{59--70}.
\bibitem[{Tse et~al.(2008)Tse, Gold, and Kidner}]{RN79}
\bibinfo{author}{R.~O.~C. Tse}, \bibinfo{author}{C.~Gold},
  \bibinfo{author}{D.~Kidner}, \bibinfo{title}{{3D} City Modelling from {LiDAR}
  Data}, \bibinfo{publisher}{Springer Berlin Heidelberg},
  \bibinfo{address}{Berlin, Heidelberg}, pp. \bibinfo{pages}{161--175}.
\bibitem[{Rabbani et~al.(2007)Rabbani, Dijkman, van~den Heuvel, and
  Vosselman}]{RN85}
\bibinfo{author}{T.~Rabbani}, \bibinfo{author}{S.~Dijkman},
  \bibinfo{author}{F.~van~den Heuvel}, \bibinfo{author}{G.~Vosselman},
\newblock \bibinfo{title}{An integrated approach for modelling and global
  registration of point clouds},
\newblock \bibinfo{journal}{ISPRS Journal of Photogrammetry and Remote Sensing}
  \bibinfo{volume}{61} (\bibinfo{year}{2007}) \bibinfo{pages}{355--370}.
\bibitem[{Biljecki et~al.(2014)Biljecki, Ledoux, Stoter, and Zhao}]{RN86}
\bibinfo{author}{F.~Biljecki}, \bibinfo{author}{H.~Ledoux},
  \bibinfo{author}{J.~Stoter}, \bibinfo{author}{J.~Zhao},
\newblock \bibinfo{title}{Formalisation of the level of detail in {3D} city
  modelling},
\newblock \bibinfo{journal}{Computers, Environment and Urban Systems}
  \bibinfo{volume}{48} (\bibinfo{year}{2014}) \bibinfo{pages}{1--15}.
\bibitem[{Gröger and Plümer(2012)}]{RN87}
\bibinfo{author}{G.~Gröger}, \bibinfo{author}{L.~Plümer},
\newblock \bibinfo{title}{{CityGML} – interoperable semantic {3D} city
  models},
\newblock \bibinfo{journal}{ISPRS Journal of Photogrammetry and Remote Sensing}
  \bibinfo{volume}{71} (\bibinfo{year}{2012}) \bibinfo{pages}{12--33}.
\bibitem[{Biljecki et~al.(2016)Biljecki, Ledoux, and Stoter}]{RN130}
\bibinfo{author}{F.~Biljecki}, \bibinfo{author}{H.~Ledoux},
  \bibinfo{author}{J.~Stoter},
\newblock \bibinfo{title}{An improved {LOD} specification for {3D} building
  models},
\newblock \bibinfo{journal}{Computers Environment and Urban Systems}
  \bibinfo{volume}{59} (\bibinfo{year}{2016}) \bibinfo{pages}{25--37}.
\bibitem[{Hongjoo~Park and Lim(2011)}]{RN321}
\bibinfo{author}{M.~S. Hongjoo~Park}, \bibinfo{author}{S.~Lim},
\newblock \bibinfo{title}{Accuracy of {3D} models derived from aerial laser
  scanning and aerial ortho-imagery},
\newblock \bibinfo{journal}{Survey Review} \bibinfo{volume}{43}
  (\bibinfo{year}{2011}) \bibinfo{pages}{109--122}.
\bibitem[{Borkowski and J\'o\'zk\'ow(2012)}]{RN320}
\bibinfo{author}{A.~Borkowski}, \bibinfo{author}{G.~J\'o\'zk\'ow},
\newblock \bibinfo{title}{Accuracy assessment of building models created from
  laser scanning data},
\newblock \bibinfo{journal}{The International Archives of the Photogrammetry,
  Remote Sensing and Spatial Information Sciences} \bibinfo{volume}{XXXIX-B3}
  (\bibinfo{year}{2012}) \bibinfo{pages}{253--258}.
\bibitem[{Chen et~al.(2017)Chen, Xue, and Lu}]{RN322}
\bibinfo{author}{K.~Chen}, \bibinfo{author}{F.~Xue}, \bibinfo{author}{W.~Lu},
\newblock \bibinfo{title}{Development of {3D} building models using
  multi-source data: A study of high-density urban area in{ Hong Kong}}
  (\bibinfo{year}{2017}) \bibinfo{pages}{609--616}.
\bibitem[{Bagheri et~al.(2018)Bagheri, Schmitt, and Zhu}]{bagheri2018fusion}
\bibinfo{author}{H.~Bagheri}, \bibinfo{author}{M.~Schmitt},
  \bibinfo{author}{X.~X. Zhu},
\newblock \bibinfo{title}{Fusion of {TanDEM-X} and {Cartosat-1} elevation data
  supported by neural network-predicted weight maps},
\newblock \bibinfo{journal}{ISPRS journal of photogrammetry and remote sensing}
  \bibinfo{volume}{144} (\bibinfo{year}{2018}) \bibinfo{pages}{285--297}.
\bibitem[{Bagheri et~al.(2017{\natexlab{a}})Bagheri, Schmitt, and
  Zhu}]{bagheri2017uncertainty}
\bibinfo{author}{H.~Bagheri}, \bibinfo{author}{M.~Schmitt},
  \bibinfo{author}{X.~X. Zhu},
\newblock \bibinfo{title}{Uncertainty assessment and weight map generation for
  efficient fusion of {TanDEM-X} and {Cartosat-1 DEMs}},
\newblock \bibinfo{journal}{The International Archives of the Photogrammetry,
  Remote Sensing and Spatial Information Sciences} \bibinfo{volume}{42}
  (\bibinfo{year}{2017}{\natexlab{a}}) \bibinfo{pages}{433--439}.
\bibitem[{Bagheri et~al.(2017{\natexlab{b}})Bagheri, Schmitt, and
  Zhu}]{bagheri2017fusion}
\bibinfo{author}{H.~Bagheri}, \bibinfo{author}{M.~Schmitt},
  \bibinfo{author}{X.~X. Zhu},
\newblock \bibinfo{title}{Fusion of {TanDEM-X} and {Cartosat-1} dems using
  {TV-norm} regularization and {ANN}-predicted weights},
\newblock in: \bibinfo{booktitle}{2017 IEEE International Geoscience and Remote
  Sensing Symposium (IGARSS)}, pp. \bibinfo{pages}{3369--3372}.
\bibitem[{Bagheri et~al.(2018)Bagheri, Schmitt, and
  Zhu}]{bagheri2018fusionIEEE}
\bibinfo{author}{H.~Bagheri}, \bibinfo{author}{M.~Schmitt},
  \bibinfo{author}{X.~X. Zhu},
\newblock \bibinfo{title}{Fusion of urban {TanDEM-X} raw {DEMs} using
  variational models},
\newblock \bibinfo{journal}{IEEE Journal of Selected Topics in Applied Earth
  Observations and Remote Sensing} \bibinfo{volume}{11} (\bibinfo{year}{2018})
  \bibinfo{pages}{4761--4774}.
\bibitem[{Bagheri et~al.(2014)Bagheri, Sadjadi, and
  Sadeghian}]{bagheri2014exploring}
\bibinfo{author}{H.~Bagheri}, \bibinfo{author}{S.~Y. Sadjadi},
  \bibinfo{author}{S.~Sadeghian},
\newblock \bibinfo{title}{Exploring the role of genetic algorithms and
  artificial neural networks for interpolation of elevation in geoinformation
  models},
\newblock \bibinfo{journal}{Innovations in 3D Geo-Information Sciences}
  (\bibinfo{year}{2014}) \bibinfo{pages}{107--121}.
\bibitem[{Bagheri et~al.(2018)Bagheri, Schmitt, d'Angelo, and
  Zhu}]{bagheri2018exploring}
\bibinfo{author}{H.~Bagheri}, \bibinfo{author}{M.~Schmitt},
  \bibinfo{author}{P.~d'Angelo}, \bibinfo{author}{X.~Zhu},
\newblock \bibinfo{title}{Exploring the applicability of semi-global matching
  for {SAR}-optical stereogrammetry of urban scenes},
\newblock \bibinfo{journal}{The International Archives of the Photogrammetry,
  Remote Sensing and Spatial Information Sciences} \bibinfo{volume}{42}
  (\bibinfo{year}{2018}) \bibinfo{pages}{43--48}.
\bibitem[{Zhang et~al.(2006)Zhang, Yan, and Chen}]{RN241}
\bibinfo{author}{K.~Zhang}, \bibinfo{author}{J.~Yan}, \bibinfo{author}{S.-C.
  Chen},
\newblock \bibinfo{title}{Automatic construction of building footprints from
  airborne {LiDAR} data},
\newblock \bibinfo{journal}{IEEE Transactions on Geoscience and Remote Sensing}
  \bibinfo{volume}{44} (\bibinfo{year}{2006}) \bibinfo{pages}{2523--2533}.
\bibitem[{Cao et~al.(2020)Cao, Weng, Du, Li, Zhong, and Mo}]{RN325}
\bibinfo{author}{S.~Cao}, \bibinfo{author}{Q.~Weng}, \bibinfo{author}{M.~Du},
  \bibinfo{author}{B.~Li}, \bibinfo{author}{R.~Zhong}, \bibinfo{author}{Y.~Mo},
\newblock \bibinfo{title}{Multi-scale three-dimensional detection of urban
  buildings using aerial lidar data},
\newblock \bibinfo{journal}{GIScience \& Remote Sensing} \bibinfo{volume}{57}
  (\bibinfo{year}{2020}) \bibinfo{pages}{1125 -- 1143}.
\bibitem[{Labetski et~al.(2023)Labetski, Vitalis, Biljecki, Ohori, and
  Stoter}]{RN326}
\bibinfo{author}{A.~Labetski}, \bibinfo{author}{S.~Vitalis},
  \bibinfo{author}{F.~Biljecki}, \bibinfo{author}{K.~A. Ohori},
  \bibinfo{author}{J.~Stoter},
\newblock \bibinfo{title}{{3D} building metrics for urban morphology},
\newblock \bibinfo{journal}{International Journal of Geographical Information
  Science} \bibinfo{volume}{37} (\bibinfo{year}{2023}) \bibinfo{pages}{36--67}.
\bibitem[{Biljecki and Chow(2022)}]{RN248}
\bibinfo{author}{F.~Biljecki}, \bibinfo{author}{Y.~S. Chow},
\newblock \bibinfo{title}{Global building morphology indicators},
\newblock \bibinfo{journal}{Computers, Environment and Urban Systems}
  \bibinfo{volume}{95} (\bibinfo{year}{2022}) \bibinfo{pages}{101809}.
\bibitem[{Huang and Zhang(2012)}]{RN332}
\bibinfo{author}{X.~Huang}, \bibinfo{author}{L.~Zhang},
\newblock \bibinfo{title}{Morphological building/shadow index for building
  extraction from high-resolution imagery over urban areas},
\newblock \bibinfo{journal}{IEEE Journal of Selected Topics in Applied Earth
  Observations and Remote Sensing} \bibinfo{volume}{5} (\bibinfo{year}{2012})
  \bibinfo{pages}{161--172}.
\bibitem[{Shi et~al.(2018)Shi, Xie, Fung, and Ng}]{RN327}
\bibinfo{author}{Y.~Shi}, \bibinfo{author}{X.~Xie}, \bibinfo{author}{J.~C.-H.
  Fung}, \bibinfo{author}{E.~Ng},
\newblock \bibinfo{title}{Identifying critical building morphological design
  factors of street-level air pollution dispersion in high-density built
  environment using mobile monitoring},
\newblock \bibinfo{journal}{Building and Environment} \bibinfo{volume}{128}
  (\bibinfo{year}{2018}) \bibinfo{pages}{248--259}.
\bibitem[{Tian et~al.(2019)Tian, Zhou, Qian, Zheng, and Yan}]{RN329}
\bibinfo{author}{Y.~Tian}, \bibinfo{author}{W.~Zhou},
  \bibinfo{author}{Y.~Qian}, \bibinfo{author}{Z.~Zheng},
  \bibinfo{author}{J.~Yan},
\newblock \bibinfo{title}{The effect of urban {2D} and {3D} morphology on air
  temperature in residential neighborhoods},
\newblock \bibinfo{journal}{Landscape Ecology} \bibinfo{volume}{34}
  (\bibinfo{year}{2019}) \bibinfo{pages}{1161--1178}.
\bibitem[{Cao et~al.(2021)Cao, Luan, Liu, and Wang}]{RN328}
\bibinfo{author}{Q.~Cao}, \bibinfo{author}{Q.~Luan}, \bibinfo{author}{Y.~Liu},
  \bibinfo{author}{R.~Wang},
\newblock \bibinfo{title}{The effects of {2D} and {3D} building morphology on
  urban environments: A multi-scale analysis in the {Beijing} metropolitan
  region},
\newblock \bibinfo{journal}{Building and Environment} \bibinfo{volume}{192}
  (\bibinfo{year}{2021}) \bibinfo{pages}{107635}.
\bibitem[{Zhang et~al.(2022)Zhang, Li, Wei, and Hu}]{RN330}
\bibinfo{author}{J.~Zhang}, \bibinfo{author}{Z.~Li}, \bibinfo{author}{Y.~Wei},
  \bibinfo{author}{D.~Hu},
\newblock \bibinfo{title}{The impact of the building morphology on microclimate
  and thermal comfort-a case study in {Beijing}},
\newblock \bibinfo{journal}{Building and Environment} \bibinfo{volume}{223}
  (\bibinfo{year}{2022}) \bibinfo{pages}{109469}.
\bibitem[{Kantianis(2022)}]{RN331}
\bibinfo{author}{D.~D. Kantianis},
\newblock \bibinfo{title}{Design morphology complexity and conceptual building
  project cost forecasting},
\newblock \bibinfo{journal}{Journal of Financial Management of Property and
  Construction} \bibinfo{volume}{27} (\bibinfo{year}{2022})
  \bibinfo{pages}{387--414}.
\bibitem[{Arefi(2009)}]{dlr60168}
\bibinfo{author}{H.~Arefi}, \bibinfo{title}{From {LiDAR} Point Clouds to 3D
  Building Models}, Ph.D. thesis, \bibinfo{year}{2009}.
\bibitem[{Macay~Moreia et~al.(2013)Macay~Moreia, Nex, Agugiaro, Remondino, and
  Lim}]{RN80}
\bibinfo{author}{J.~M. Macay~Moreia}, \bibinfo{author}{F.~Nex},
  \bibinfo{author}{G.~Agugiaro}, \bibinfo{author}{F.~Remondino},
  \bibinfo{author}{N.~J. Lim},
\newblock \bibinfo{title}{From {DSM} to {3D} building models: A quantitative
  evaluation},
\newblock \bibinfo{journal}{Int. Arch. Photogramm. Remote Sens. Spatial Inf.
  Sci.} \bibinfo{volume}{XL-1/W1} (\bibinfo{year}{2013})
  \bibinfo{pages}{213--219}. \bibinfo{note}{ISPRS-Archives}.
\bibitem[{Rajpriya et~al.(2014)Rajpriya, Vyas, and Sharma}]{RN36}
\bibinfo{author}{N.~R. Rajpriya}, \bibinfo{author}{A.~Vyas},
  \bibinfo{author}{S.~A. Sharma},
\newblock \bibinfo{title}{Generation of {3D} model for urban area using
  {Ikonos} and {Cartosat-1} satellite imageries with {RS} and {GIS}
  techniques},
\newblock \bibinfo{journal}{Int. Arch. Photogramm. Remote Sens. Spatial Inf.
  Sci.} \bibinfo{volume}{XL-8} (\bibinfo{year}{2014})
  \bibinfo{pages}{899--906}. \bibinfo{note}{ISPRS-Archives}.
\bibitem[{Park and Guldmann(2019)}]{RN81}
\bibinfo{author}{Y.~Park}, \bibinfo{author}{J.-M. Guldmann},
\newblock \bibinfo{title}{Creating {3D} city models with building footprints
  and {LiDAR} point cloud classification: A machine learning approach},
\newblock \bibinfo{journal}{Computers, Environment and Urban Systems}
  \bibinfo{volume}{75} (\bibinfo{year}{2019}) \bibinfo{pages}{76--89}.
\bibitem[{Gruen et~al.(2019)Gruen, Schubiger, Qin, Schrotter, Xiong, Li, Ling,
  Xiao, Yao, and Nuesch}]{RN82}
\bibinfo{author}{A.~Gruen}, \bibinfo{author}{S.~Schubiger},
  \bibinfo{author}{R.~Qin}, \bibinfo{author}{G.~Schrotter},
  \bibinfo{author}{B.~Xiong}, \bibinfo{author}{J.~Li},
  \bibinfo{author}{X.~Ling}, \bibinfo{author}{C.~Xiao},
  \bibinfo{author}{S.~Yao}, \bibinfo{author}{F.~Nuesch},
\newblock \bibinfo{title}{Semantically enriched high resolution {LOD3} building
  model generation},
\newblock \bibinfo{journal}{Int. Arch. Photogramm. Remote Sens. Spatial Inf.
  Sci.} \bibinfo{volume}{XLII-4/W15} (\bibinfo{year}{2019})
  \bibinfo{pages}{11--18}. \bibinfo{note}{ISPRS-Archives}.
\bibitem[{Wen et~al.(2019)Wen, Xie, Liu, and Yan}]{RN96}
\bibinfo{author}{X.~Wen}, \bibinfo{author}{H.~Xie}, \bibinfo{author}{H.~Liu},
  \bibinfo{author}{L.~Yan},
\newblock \bibinfo{title}{Accurate reconstruction of the {LOD3} building model
  by integrating multi-source point clouds and oblique remote sensing imagery},
\newblock \bibinfo{journal}{ISPRS International Journal of Geo-Information}
  \bibinfo{volume}{8} (\bibinfo{year}{2019}) \bibinfo{pages}{135}.
\bibitem[{Partovi et~al.(2019)Partovi, Fraundorfer, Bahmanyar, Huang, and
  Reinartz}]{RN35}
\bibinfo{author}{T.~Partovi}, \bibinfo{author}{F.~Fraundorfer},
  \bibinfo{author}{R.~Bahmanyar}, \bibinfo{author}{H.~Huang},
  \bibinfo{author}{P.~Reinartz},
\newblock \bibinfo{title}{Automatic {3D} building model reconstruction from
  very high resolution stereo satellite imagery},
\newblock \bibinfo{journal}{Remote Sensing} \bibinfo{volume}{11}
  (\bibinfo{year}{2019}) \bibinfo{pages}{1660}.
\bibitem[{Bagheri et~al.(2019)Bagheri, Schmitt, and Zhu}]{baghfusion}
\bibinfo{author}{H.~Bagheri}, \bibinfo{author}{M.~Schmitt},
  \bibinfo{author}{X.~Zhu},
\newblock \bibinfo{title}{Fusion of multi-sensor-derived heights and
  {OSM}-derived building footprints for urban {3D} reconstruction},
\newblock \bibinfo{journal}{ISPRS International Journal of Geo-Information}
  \bibinfo{volume}{8} (\bibinfo{year}{2019}) \bibinfo{pages}{193}.
\bibitem[{Buyuksalih et~al.(2019)Buyuksalih, Baskaraca, Bayburt, Buyuksalih,
  and Abdul~Rahman}]{RN240}
\bibinfo{author}{G.~Buyuksalih}, \bibinfo{author}{P.~Baskaraca},
  \bibinfo{author}{S.~Bayburt}, \bibinfo{author}{I.~Buyuksalih},
  \bibinfo{author}{A.~Abdul~Rahman},
\newblock \bibinfo{title}{{3D} city modelling of {Istanbul} based on {LiDAR}
  data and panoramic images – issues and challenges},
\newblock \bibinfo{journal}{The International Archives of the Photogrammetry,
  Remote Sensing and Spatial Information Sciences} \bibinfo{volume}{XLII-4/W12}
  (\bibinfo{year}{2019}) \bibinfo{pages}{51--60}.
\bibitem[{Pepe et~al.(2021)Pepe, Costantino, Alfio, Vozza, and
  Cartellino}]{RN34}
\bibinfo{author}{M.~Pepe}, \bibinfo{author}{D.~Costantino},
  \bibinfo{author}{V.~S. Alfio}, \bibinfo{author}{G.~Vozza},
  \bibinfo{author}{E.~Cartellino},
\newblock \bibinfo{title}{A novel method based on deep learning, {GIS} and
  geomatics software for building a {3D} city model from {VHR} satellite stereo
  imagery},
\newblock \bibinfo{journal}{ISPRS International Journal of Geo-Information}
  \bibinfo{volume}{10} (\bibinfo{year}{2021}) \bibinfo{pages}{697}.
\bibitem[{Huang et~al.(2022)Huang, Stoter, Peters, and Nan}]{rs14092254}
\bibinfo{author}{J.~Huang}, \bibinfo{author}{J.~Stoter},
  \bibinfo{author}{R.~Peters}, \bibinfo{author}{L.~Nan},
\newblock \bibinfo{title}{{City3D}: Large-scale building reconstruction from
  airborne {LiDAR} point clouds},
\newblock \bibinfo{journal}{Remote Sensing} \bibinfo{volume}{14}
  (\bibinfo{year}{2022}).
\bibitem[{Wang et~al.(2023)Wang, Zhou, Hu, Wang, Fu, Li, and Xie}]{RN323}
\bibinfo{author}{F.~Wang}, \bibinfo{author}{G.~Zhou}, \bibinfo{author}{H.~Hu},
  \bibinfo{author}{Y.~Wang}, \bibinfo{author}{B.~Fu}, \bibinfo{author}{S.~Li},
  \bibinfo{author}{J.~Xie},
\newblock \bibinfo{title}{Reconstruction of {LOD2} building models guided by
  façade structures from oblique photogrammetric point cloud},
\newblock \bibinfo{journal}{Remote Sensing} \bibinfo{volume}{15}
  (\bibinfo{year}{2023}).
\bibitem[{Mao et~al.(2023)Mao, Chen, Zhao, Chen, Tang, Liu, Wang, Diao, Sun,
  and Fu}]{RN324}
\bibinfo{author}{Y.~Mao}, \bibinfo{author}{K.~Chen}, \bibinfo{author}{L.~Zhao},
  \bibinfo{author}{W.~Chen}, \bibinfo{author}{D.~Tang},
  \bibinfo{author}{W.~Liu}, \bibinfo{author}{Z.~Wang},
  \bibinfo{author}{W.~Diao}, \bibinfo{author}{X.~Sun}, \bibinfo{author}{K.~Fu},
\newblock \bibinfo{title}{Elevation estimation-driven building {3D}
  reconstruction from single-view remote sensing imagery},
\newblock \bibinfo{journal}{IEEE Transactions on Geoscience and Remote Sensing}
  \bibinfo{volume}{61} (\bibinfo{year}{2023}) \bibinfo{pages}{1--18}.
\bibitem[{ANH(2023)}]{RN123}
\bibinfo{author}{ANH}, \bibinfo{title}{{AHN} quality description},
  \bibinfo{howpublished}{\url{https://www.ahn.nl/kwaliteitsbeschrijving}},
  \bibinfo{year}{2023}. \bibinfo{note}{Accessed: Sep 02, 2023}.
\bibitem[{USGS(2023)}]{RN221}
\bibinfo{author}{USGS}, \bibinfo{title}{{2018 {USGS}/{NRCS} {LiDAR} {DEM}:
  Southeast {Florida}}},
  \bibinfo{howpublished}{\url{https://www.fisheries.noaa.gov/inport/item/59009}},
  \bibinfo{year}{2023}. \bibinfo{note}{[Accessed: Sep 02, 2023]}.
\bibitem[{Pan and Yang(2010)}]{RN129}
\bibinfo{author}{S.~J. Pan}, \bibinfo{author}{Q.~Yang},
\newblock \bibinfo{title}{A survey on transfer learning},
\newblock \bibinfo{journal}{IEEE Transactions on Knowledge and Data
  Engineering} \bibinfo{volume}{22} (\bibinfo{year}{2010})
  \bibinfo{pages}{1345--1359}.
\bibitem[{Hecht et~al.(2013)Hecht, Kunze, and Hahmann}]{RN190}
\bibinfo{author}{R.~Hecht}, \bibinfo{author}{C.~Kunze},
  \bibinfo{author}{S.~Hahmann},
\newblock \bibinfo{title}{Measuring completeness of building footprints in
  {OpenStreetMap} over space and time},
\newblock \bibinfo{journal}{ISPRS International Journal of Geo-Information}
  \bibinfo{volume}{2} (\bibinfo{year}{2013}) \bibinfo{pages}{1066--1091}.
\bibitem[{Zhuo et~al.(2018)Zhuo, Fraundorfer, Kurz, and Reinartz}]{RN147}
\bibinfo{author}{X.~Zhuo}, \bibinfo{author}{F.~Fraundorfer},
  \bibinfo{author}{F.~Kurz}, \bibinfo{author}{P.~Reinartz},
\newblock \bibinfo{title}{Optimization of {OpenStreetMap} building footprints
  based on semantic information of oblique {UAV} images},
\newblock \bibinfo{journal}{Remote Sensing} \bibinfo{volume}{10}
  (\bibinfo{year}{2018}) \bibinfo{pages}{624}.
\bibitem[{Brovelli and Zamboni(2018)}]{RN189}
\bibinfo{author}{M.~A. Brovelli}, \bibinfo{author}{G.~Zamboni},
\newblock \bibinfo{title}{A new method for the assessment of spatial accuracy
  and completeness of {OpenStreetMap} building footprints},
\newblock \bibinfo{journal}{ISPRS International Journal of Geo-Information}
  \bibinfo{volume}{7} (\bibinfo{year}{2018}) \bibinfo{pages}{289}.
\bibitem[{Gribov(2019)}]{RN228}
\bibinfo{author}{A.~Gribov},
\newblock \bibinfo{title}{Optimal compression of a polyline while aligning to
  preferred directions},
\newblock in: \bibinfo{booktitle}{2019 International Conference on Document
  Analysis and Recognition Workshops (ICDARW)}, volume~\bibinfo{volume}{1}, pp.
  \bibinfo{pages}{98--102}.
\bibitem[{Arefi et~al.(2008)Arefi, Engels, Hahn, and Mayer}]{RN238}
\bibinfo{author}{H.~Arefi}, \bibinfo{author}{J.~Engels},
  \bibinfo{author}{M.~Hahn}, \bibinfo{author}{H.~Mayer},
\newblock \bibinfo{title}{Levels of detail in {3D} building reconstruction from
  {LiDAR} data},
\newblock \bibinfo{journal}{Proceedings of the International Archieves If the
  Photogrammetry, Remote Sensing, and Spatial Information Sciences}
  \bibinfo{volume}{37} (\bibinfo{year}{2008}) \bibinfo{pages}{485--490}.
\bibitem[{Stoter et~al.(2014)Stoter, Vosselman, Dahmen, Oude~Elberink, and
  Ledoux}]{RN239}
\bibinfo{author}{J.~Stoter}, \bibinfo{author}{G.~Vosselman},
  \bibinfo{author}{C.~Dahmen}, \bibinfo{author}{S.~Oude~Elberink},
  \bibinfo{author}{H.~Ledoux},
\newblock \bibinfo{title}{{CityGML} implementation specifications for a
  countrywide {3D} data set: The case of the {Netherlands}},
\newblock \bibinfo{journal}{Photogrammetric Engineering \& Remote Sensing}
  \bibinfo{volume}{80} (\bibinfo{year}{2014}) \bibinfo{pages}{1069--1077}.
\bibitem[{Biljecki et~al.(2016)Biljecki, Ledoux, Stoter, and Vosselman}]{RN237}
\bibinfo{author}{F.~Biljecki}, \bibinfo{author}{H.~Ledoux},
  \bibinfo{author}{J.~Stoter}, \bibinfo{author}{G.~Vosselman},
\newblock \bibinfo{title}{The variants of an {LOD} of a {3D} building model and
  their influence on spatial analyses},
\newblock \bibinfo{journal}{ISPRS Journal of Photogrammetry and Remote Sensing}
  \bibinfo{volume}{116} (\bibinfo{year}{2016}) \bibinfo{pages}{42--54}.
\bibitem[{Everingham et~al.(2015)Everingham, Eslami, Van~Gool, Williams, Winn,
  and Zisserman}]{RN131}
\bibinfo{author}{M.~Everingham}, \bibinfo{author}{S.~M.~A. Eslami},
  \bibinfo{author}{L.~Van~Gool}, \bibinfo{author}{C.~K.~I. Williams},
  \bibinfo{author}{J.~Winn}, \bibinfo{author}{A.~Zisserman},
\newblock \bibinfo{title}{The pascal visual object classes challenge: A
  retrospective},
\newblock \bibinfo{journal}{International Journal of Computer Vision}
  \bibinfo{volume}{111} (\bibinfo{year}{2015}) \bibinfo{pages}{98--136}.
\bibitem[{Chawla et~al.(2002)Chawla, Bowyer, Hall, and Kegelmeyer}]{RN113}
\bibinfo{author}{N.~Chawla}, \bibinfo{author}{K.~Bowyer},
  \bibinfo{author}{L.~Hall}, \bibinfo{author}{W.~Kegelmeyer},
\newblock \bibinfo{title}{{SMOTE}: Synthetic minority over-sampling technique},
\newblock \bibinfo{journal}{J. Artif. Intell. Res. (JAIR)} \bibinfo{volume}{16}
  (\bibinfo{year}{2002}) \bibinfo{pages}{321--357}.
\bibitem[{Taha and Hanbury(2015)}]{RN132}
\bibinfo{author}{A.~A. Taha}, \bibinfo{author}{A.~Hanbury},
\newblock \bibinfo{title}{Metrics for evaluating {3D} medical image
  segmentation: analysis, selection, and tool},
\newblock \bibinfo{journal}{BMC Medical Imaging} \bibinfo{volume}{15}
  (\bibinfo{year}{2015}) \bibinfo{pages}{29}.
\bibitem[{Teo(2019)}]{RN343}
\bibinfo{author}{T.-A. Teo},
\newblock \bibinfo{title}{Deep-learning for {LOD1} building reconstruction from
  airborne {LiDAR} data},
\newblock in: \bibinfo{booktitle}{IGARSS 2019 - 2019 IEEE International
  Geoscience and Remote Sensing Symposium}, pp. \bibinfo{pages}{86--89}.
\bibitem[{Bool et~al.(2018)Bool, Mabaquiao, Tupas, and Fabila}]{RN349}
\bibinfo{author}{D.~L. Bool}, \bibinfo{author}{L.~C. Mabaquiao},
  \bibinfo{author}{M.~E. Tupas}, \bibinfo{author}{J.~L. Fabila},
\newblock \bibinfo{title}{Automated building detection using ransac from
  classified lidar point cloud data},
\newblock \bibinfo{journal}{The International Archives of the Photogrammetry,
  Remote Sensing and Spatial Information Sciences} \bibinfo{volume}{XLII-4/W9}
  (\bibinfo{year}{2018}) \bibinfo{pages}{115--121}.
\bibitem[{Gamal et~al.(2020)Gamal, Wibisono, Wicaksono, Abyan, Hamid, Wisesa,
  Jatmiko, and Ardhianto}]{RN341}
\bibinfo{author}{A.~Gamal}, \bibinfo{author}{A.~Wibisono},
  \bibinfo{author}{S.~Wicaksono}, \bibinfo{author}{M.~Abyan},
  \bibinfo{author}{N.~Hamid}, \bibinfo{author}{H.~Wisesa},
  \bibinfo{author}{W.~Jatmiko}, \bibinfo{author}{R.~Ardhianto},
\newblock \bibinfo{title}{Automatic {LiDAR} building segmentation based on
  {DGCNN} and euclidean clustering},
\newblock \bibinfo{journal}{Journal of Big Data} \bibinfo{volume}{7}
  (\bibinfo{year}{2020}).
\bibitem[{Yang et~al.(2024)Yang, Zhang, Liu, and Gao}]{RN351}
\bibinfo{author}{W.~Yang}, \bibinfo{author}{Y.~Zhang},
  \bibinfo{author}{X.~Liu}, \bibinfo{author}{B.~Gao},
\newblock \bibinfo{title}{Scene adaptive building individual segmentation based
  on large-scale airborne lidar point clouds},
\newblock \bibinfo{journal}{IEEE Transactions on Geoscience and Remote Sensing}
  \bibinfo{volume}{62} (\bibinfo{year}{2024}) \bibinfo{pages}{1--15}.
\bibitem[{Cheng et~al.(2011)Cheng, Gong, Li, and Liu}]{RN350}
\bibinfo{author}{L.~Cheng}, \bibinfo{author}{J.~Gong}, \bibinfo{author}{M.~Li},
  \bibinfo{author}{Y.~Liu},
\newblock \bibinfo{title}{{3D} building model reconstruction from multi-view
  aerial imagery and lidar data},
\newblock \bibinfo{journal}{Photogrammetric Engineering and Remote Sensing}
  \bibinfo{volume}{77} (\bibinfo{year}{2011}) \bibinfo{pages}{125--139}.
\bibitem[{Sulaiman et~al.(2024)Sulaiman, Finnesand, Farmanbar, Belbachir, and
  Rong}]{RN342}
\bibinfo{author}{M.~Sulaiman}, \bibinfo{author}{E.~Finnesand},
  \bibinfo{author}{M.~Farmanbar}, \bibinfo{author}{A.~N. Belbachir},
  \bibinfo{author}{C.~Rong},
\newblock \bibinfo{title}{Building precision: Efficient {Encoder–Decoder}
  networks for remote sensing based on aerial {RGB} and {LiDAR} data},
\newblock \bibinfo{journal}{IEEE Access} \bibinfo{volume}{12}
  (\bibinfo{year}{2024}) \bibinfo{pages}{60329--60346}.
\bibitem[{Bittner et~al.(2018)Bittner, Adam, Cui, Körner, and
  Reinartz}]{RN344}
\bibinfo{author}{K.~Bittner}, \bibinfo{author}{F.~Adam},
  \bibinfo{author}{S.~Cui}, \bibinfo{author}{M.~Körner},
  \bibinfo{author}{P.~Reinartz},
\newblock \bibinfo{title}{Building footprint extraction from {VHR} remote
  sensing images combined with normalized {DSMs} using fused fully
  convolutional networks},
\newblock \bibinfo{journal}{IEEE Journal of Selected Topics in Applied Earth
  Observations and Remote Sensing} \bibinfo{volume}{11} (\bibinfo{year}{2018})
  \bibinfo{pages}{2615--2629}.
\bibitem[{Vulkan et~al.(2018)Vulkan, Kloog, Dorman, and Erell}]{RN345}
\bibinfo{author}{A.~Vulkan}, \bibinfo{author}{I.~Kloog},
  \bibinfo{author}{M.~Dorman}, \bibinfo{author}{E.~Erell},
\newblock \bibinfo{title}{Modeling the potential for pv installation in
  residential buildings in dense urban areas},
\newblock \bibinfo{journal}{Energy and Buildings} \bibinfo{volume}{169}
  (\bibinfo{year}{2018}) \bibinfo{pages}{97--109}.
\bibitem[{Hosseini and Bagheri(ress)}]{Hosseini_undated-qc}
\bibinfo{author}{M.~Hosseini}, \bibinfo{author}{H.~Bagheri},
\newblock \bibinfo{title}{Improving the resolution of solar energy potential
  maps derived from global {DSMs} for rooftop solar panel placement using deep
  learning},
\newblock \bibinfo{journal}{Heliyon}  (\bibinfo{year}{In press}).
\bibitem[{Ranjbar et~al.(2012)Ranjbar, Gharagozlou, and Nejad}]{RN346}
\bibinfo{author}{H.~R. Ranjbar}, \bibinfo{author}{A.~R. Gharagozlou},
  \bibinfo{author}{A.~R.~V. Nejad},
\newblock \bibinfo{title}{3d analysis and investigation of traffic noise impact
  from hemmat highway located in {Tehran} on buildings and surrounding areas}
  (\bibinfo{year}{2012}).
\bibitem[{Biljecki et~al.(2017)Biljecki, Ledoux, and Stoter}]{RN347}
\bibinfo{author}{F.~Biljecki}, \bibinfo{author}{H.~Ledoux},
  \bibinfo{author}{J.~Stoter}, \bibinfo{title}{Does a Finer Level of Detail of
  a {3D} City Model Bring an Improvement for Estimating Shadows?}, pp.
  \bibinfo{pages}{31--47}.
\bibitem[{Zhou et~al.(2022)Zhou, Yuan, Hu, Wei, Dang, and Sun}]{RN348}
\bibinfo{author}{L.~Zhou}, \bibinfo{author}{B.~Yuan}, \bibinfo{author}{F.~Hu},
  \bibinfo{author}{C.~Wei}, \bibinfo{author}{X.~Dang},
  \bibinfo{author}{D.~Sun},
\newblock \bibinfo{title}{Understanding the effects of {2D/3D} urban morphology
  on land surface temperature based on local climate zones},
\newblock \bibinfo{journal}{Building and Environment} \bibinfo{volume}{208}
  (\bibinfo{year}{2022}) \bibinfo{pages}{108578}.
\bibitem[{Ronneberger et~al.(2015)Ronneberger, Fischer, and Brox}]{RN110}
\bibinfo{author}{O.~Ronneberger}, \bibinfo{author}{P.~Fischer},
  \bibinfo{author}{T.~Brox},
\newblock \bibinfo{title}{{U-Net}: Convolutional networks for biomedical image
  segmentation},
\newblock in: \bibinfo{editor}{N.~Navab}, \bibinfo{editor}{J.~Hornegger},
  \bibinfo{editor}{W.~M. Wells}, \bibinfo{editor}{A.~F. Frangi} (Eds.),
  \bibinfo{booktitle}{Medical Image Computing and Computer-Assisted
  Intervention – {MICCAI} 2015}, \bibinfo{publisher}{Springer International
  Publishing}, \bibinfo{year}{2015}, pp. \bibinfo{pages}{234--241}.
\bibitem[{Shen et~al.(2017)Shen, Zhou, Long, Jiang, Pan, and Zhang}]{RN126}
\bibinfo{author}{T.~Shen}, \bibinfo{author}{T.~Zhou},
  \bibinfo{author}{G.~Long}, \bibinfo{author}{J.~Jiang},
  \bibinfo{author}{S.~Pan}, \bibinfo{author}{C.~Zhang},
\newblock \bibinfo{title}{{DiSAN}: Directional self-attention network for
  {RNN/CNN}-free language understanding},
\newblock \bibinfo{journal}{Proceedings of the AAAI Conference on Artificial
  Intelligence} \bibinfo{volume}{32} (\bibinfo{year}{2017}).
\bibitem[{Oktay et~al.(2018)Oktay, Schlemper, Folgoc, Lee, Heinrich, Misawa,
  Mori, McDonagh, Hammerla, Kainz, Glocker, and Rueckert}]{RN112}
\bibinfo{author}{O.~Oktay}, \bibinfo{author}{J.~Schlemper},
  \bibinfo{author}{L.~L. Folgoc}, \bibinfo{author}{M.~Lee},
  \bibinfo{author}{M.~Heinrich}, \bibinfo{author}{K.~Misawa},
  \bibinfo{author}{K.~Mori}, \bibinfo{author}{S.~McDonagh},
  \bibinfo{author}{N.~Y. Hammerla}, \bibinfo{author}{B.~Kainz},
  \bibinfo{author}{B.~Glocker}, \bibinfo{author}{D.~Rueckert},
\newblock \bibinfo{title}{Attention {U-Net}: Learning where to look for the
  pancreas},
\newblock in: \bibinfo{booktitle}{Medical Imaging with Deep Learning}.
\bibitem[{Huang et~al.(2020)Huang, Lin, Tong, Hu, Zhang, Iwamoto, Han, Chen,
  and Wu}]{RN111}
\bibinfo{author}{H.~Huang}, \bibinfo{author}{L.~Lin},
  \bibinfo{author}{R.~Tong}, \bibinfo{author}{H.~Hu},
  \bibinfo{author}{Q.~Zhang}, \bibinfo{author}{Y.~Iwamoto},
  \bibinfo{author}{X.~Han}, \bibinfo{author}{Y.-W. Chen},
  \bibinfo{author}{J.~Wu},
\newblock \bibinfo{title}{{Unet} 3+: A full-scale connected {UNet} for medical
  image segmentation},
\newblock in: \bibinfo{booktitle}{ICASSP 2020-2020 IEEE international
  conference on acoustics, speech and signal processing (ICASSP)},
  \bibinfo{publisher}{IEEE}, \bibinfo{year}{2020}, pp.
  \bibinfo{pages}{1055--1059}.
\bibitem[{Abraham and Khan(2019)}]{RN244}
\bibinfo{author}{N.~Abraham}, \bibinfo{author}{N.~M. Khan},
\newblock \bibinfo{title}{A novel focal tversky loss function with improved
  {Attention U-Net} for lesion segmentation},
\newblock in: \bibinfo{booktitle}{2019 IEEE 16th International Symposium on
  Biomedical Imaging (ISBI 2019)}, pp. \bibinfo{pages}{683--687}.
\bibitem[{Rahman and Wang(2016)}]{RN245}
\bibinfo{author}{M.~A. Rahman}, \bibinfo{author}{Y.~Wang},
\newblock \bibinfo{title}{Optimizing intersection-over-union in deep neural
  networks for image segmentation},
\newblock in: \bibinfo{editor}{G.~Bebis}, \bibinfo{editor}{R.~Boyle},
  \bibinfo{editor}{B.~Parvin}, \bibinfo{editor}{D.~Koracin},
  \bibinfo{editor}{F.~Porikli}, \bibinfo{editor}{S.~Skaff},
  \bibinfo{editor}{A.~Entezari}, \bibinfo{editor}{J.~Min},
  \bibinfo{editor}{D.~Iwai}, \bibinfo{editor}{A.~Sadagic},
  \bibinfo{editor}{C.~Scheidegger}, \bibinfo{editor}{T.~Isenberg} (Eds.),
  \bibinfo{booktitle}{Advances in Visual Computing},
  \bibinfo{publisher}{Springer International Publishing}, \bibinfo{year}{2016},
  pp. \bibinfo{pages}{234--244}.
\bibitem[{Chen et~al.(2018)Chen, Zhu, Papandreou, Schroff, and Adam}]{RN148}
\bibinfo{author}{L.-C. Chen}, \bibinfo{author}{Y.~Zhu},
  \bibinfo{author}{G.~Papandreou}, \bibinfo{author}{F.~Schroff},
  \bibinfo{author}{H.~Adam},
\newblock \bibinfo{title}{Encoder-decoder with atrous separable convolution for
  semantic image segmentation},
\newblock in: \bibinfo{booktitle}{Proceedings of the European conference on
  computer vision (ECCV)}, pp. \bibinfo{pages}{801--818}.
\bibitem[{He et~al.(2015)He, Zhang, Ren, and Sun}]{RN149}
\bibinfo{author}{K.~He}, \bibinfo{author}{X.~Zhang}, \bibinfo{author}{S.~Ren},
  \bibinfo{author}{J.~Sun},
\newblock \bibinfo{title}{Spatial pyramid pooling in deep convolutional
  networks for visual recognition},
\newblock \bibinfo{journal}{IEEE transactions on pattern analysis and machine
  intelligence} \bibinfo{volume}{37} (\bibinfo{year}{2015})
  \bibinfo{pages}{1904--1916}.
\bibitem[{Badrinarayanan et~al.(2017)Badrinarayanan, Kendall, and
  Cipolla}]{RN151}
\bibinfo{author}{V.~Badrinarayanan}, \bibinfo{author}{A.~Kendall},
  \bibinfo{author}{R.~Cipolla},
\newblock \bibinfo{title}{{Segnet}: A deep convolutional encoder-decoder
  architecture for image segmentation},
\newblock \bibinfo{journal}{IEEE transactions on pattern analysis and machine
  intelligence} \bibinfo{volume}{39} (\bibinfo{year}{2017})
  \bibinfo{pages}{2481--2495}.
\bibitem[{Chen et~al.(2017{\natexlab{a}})Chen, Papandreou, Schroff, and
  Adam}]{RN152}
\bibinfo{author}{L.-C. Chen}, \bibinfo{author}{G.~Papandreou},
  \bibinfo{author}{F.~Schroff}, \bibinfo{author}{H.~Adam},
\newblock \bibinfo{title}{Rethinking atrous convolution for semantic image
  segmentation},
\newblock \bibinfo{journal}{arXiv preprint arXiv:1706.05587}
  (\bibinfo{year}{2017}{\natexlab{a}}).
\bibitem[{Chen et~al.(2017{\natexlab{b}})Chen, Papandreou, Kokkinos, Murphy,
  and Yuille}]{RN214}
\bibinfo{author}{L.-C. Chen}, \bibinfo{author}{G.~Papandreou},
  \bibinfo{author}{I.~Kokkinos}, \bibinfo{author}{K.~Murphy},
  \bibinfo{author}{A.~L. Yuille},
\newblock \bibinfo{title}{{Deeplab}: Semantic image segmentation with deep
  convolutional nets, atrous convolution, and fully connected {CRFs}},
\newblock \bibinfo{journal}{IEEE Transactions on Pattern Analysis and Machine
  Intelligence} \bibinfo{volume}{40} (\bibinfo{year}{2017}{\natexlab{b}})
  \bibinfo{pages}{834--848}.

\end{thebibliography}
	
	\begin{appendices}
		
		\counterwithin{figure}{section}
		\counterwithin{table}{section}
		\section{Deep Segmentation Models}\label{appendix_dl}
		\begin{itemize}
			\item{U-Net}
		\end{itemize}
		One of the most well-known semantic image segmentation models is the U-Net architecture, first proposed for medical image segmentation \citep{RN110}. This model consists of convolutional neural networks (CNNs) arranged in an encoder-decoder structure. The encoder reduces image dimensions and extracts contextual information while the decoder reconstructs the segmented image. One of the notable features of this model is the use of skip connections between the encoder and decoder layers, which helps to preserve the spatial information and enhance the segmentation accuracy. At each decoder stage, these connections transfer spatial features from the encoder layers to the decoder layers, providing sufficient content for generating the segmentation mask. Due to its U-shaped structure, the model is named U-Net.
		
		\begin{itemize}
			\item{Attention U-Net}
		\end{itemize}
		
		The Attention U-Net model is a variant of the U-Net model that enhances its performance by incorporating attention mechanisms \citep{RN126}. These attention mechanisms allow the network to focus on the relevant image features important for segmentation while reducing the importance of irrelevant features \citep{RN112}. The skip connections between the encoder and decoder use the attention gate mechanism, emphasizing significant features based on attention coefficients. These coefficients, computed using gating signals from deeper layers and spatial information from skip connections, guide the network to prioritize important features during decoding.

			
			\begin{itemize}
				\item{U-Net3+}
			\end{itemize}
			The U-Net3+ model is another variant of the U-Net architecture, designed to enhance accuracy while reducing the number of network parameters \citep{RN111}. This model leverages full-scale skip connections to incorporate low- and high-detail feature maps in segmentation. Additionally, by applying full-scale deep supervision, U-Net3+ outperforms the original U-Net. U-Net3+ uses full-scale skip connections at each decoder layer, allowing the model to receive both same- and lower-scale feature maps from the encoder and larger-scale feature maps from the decoder. These connections allow the model to capture fine-grained and coarse-grained semantic details at different scales. Furthermore, applying the hybrid loss function and the classification-guided module (CGM) enhances the segmentation process and reduces the false-positive rate \citep{RN244, RN245}.
			
			
			
			\begin{itemize}
				\item{DeepLabV3+}
			\end{itemize}
			
			DeepLabV3+ represents the most recent advancement in the DeepLab semantic segmentation models developed by Google AI \citep{RN148}. By combining two key elements, the spatial pyramid pooling (SPP) module and the encoder-decoder structure, this model achieves significant accuracy and efficiency in extracting the boundaries of objects. The SPP module introduces multi-scale contextual information, improving the detection of objects at different levels of detail \citep{RN149}. The encoder-decoder structure, similar to U-Net, reduces the resolution of the input image in the encoder path to extract high-level features and restores spatial features in the decoder path \citep{RN110, RN151}. This encoder path is designed based on the DeepLabV3 architecture, with the difference that in DeepLabV3+, a simple and effective decoding module is added to the architecture to recover object boundaries precisely \citep{RN152}.
			
			In the DeepLabV3+ architecture, the resolution of the features in the encoder is controlled using atrous (dilated) convolutions, balancing model accuracy and runtime efficiency. Atrous convolutions expand the receptive field of the filters without increasing the number of parameters or computational complexity \citep{RN152, RN214}. The atrous spatial pyramid pooling (ASPP) module refines the feature map using multi-scale information through parallel atrous convolution layers.
			
			For this study, the DeepLabV3+ model utilizes ResNet50, a deep network pre-trained on ImageNet, as its backbone. Since the input images in this research were single-channel, they were converted into three-channel images to align with the model’s requirements.
		
	\end{appendices}

\end{document}